\documentclass[useAMS,usenatbib,usegraphicx,twocolumn]{mn2e}
\usepackage[figuresright]{rotating}
\usepackage{float}
\usepackage{amsmath,amssymb}

\title[Asymptotic Orbits in Barred Spiral Galaxies]{Asymptotic Orbits in Barred Spiral Galaxies}
\author[M. Harsoula, C. Kalapotharakos and G. Contopoulos]
       {M. ~Harsoula C. ~Kalapotharakos and G. ~Contopoulos\\
        Research Center for Astronomy,
           Academy of Athens, Soranou Efesiou 4, GR-115 27 Athens, Greece\\
           e-mail: mharsoul@academyofathens.gr,
           ckalapot@phys.uoa.gr, gcontop@academyofathens.gr}
\date{Released 2007 April 12}
\pagerange{\pageref{firstpage}--\pageref{lastpage}} \pubyear{2007}

\def\LaTeX{L\kern-.36em\raise.3ex\hbox{a}\kern-.15em
    T\kern-.1667em\lower.7ex\hbox{E}\kern-.125emX}

\begin{document}

\maketitle

\label{firstpage}

\begin{abstract}
We study the formation of the spiral structure of barred spiral
galaxies, using an $N$-body model. The evolution of this $N$-body
model in the adiabatic approximation maintains a strong spiral
pattern for more than 10 bar rotations. We find that this longevity
of the spiral arms is mainly due to the phenomenon of stickiness of
chaotic orbits close to the unstable asymptotic manifolds originated
from the main unstable periodic orbits, both inside and outside
corotation. The stickiness along the manifolds corresponding to
different energy levels supports parts of the spiral structure. The
loci of the disc velocity minima (where the particles spend most of
their time, in the configuration space) reveal the density maxima
and therefore the main morphological structures of the system. We
study the relation of these loci with those of the apocentres and
pericentres at different energy levels. The diffusion of the sticky
chaotic orbits outwards is slow and depends on the initial
conditions and the corresponding Jacobi constant.
\end{abstract}


\section{Introduction}

Galactic dynamics together with celestial mechanics have played a
leading role in the study of orbits as well as in the study of order
and chaos. Many decades ago and for a long time the main interest
was about regular orbits since they were considered as the main
building blocks of galaxies. The main idea was that regular orbits
are structurally robust and therefore they are able to support
various morphological features. The most extreme manifestation was
elliptical galaxies for which no chaotic orbits were considered to
exist \citep{b30,b31}. In rotating galaxies the picture was more
confused, especially after the study of the main periodic orbits in
systems representing barred galaxies. These studies showed the
existence of many unstable orbits (especially near corotation) which
are generators of chaos \citep{b32}. Chaos was found in the 80's
near the ends of the bar \citep{b16,b33}. However, chaos was not
considered as responsible for some of the main features of the
galaxies like the spiral arms beyond corotation.

The importance of chaos was realized mainly in the 90's, with
application both in elliptical and in barred spiral galaxies, mostly
in steady state cases \citep{b34,b17}.

Only in the 2000's people started to consider asymptotic orbits
emanating from the unstable periodic orbits in explaining the main
features of galaxies, like the spiral structure in barred spiral
galaxies. The asymptotic orbits have initial conditions on the
asymptotic manifolds of unstable periodic orbits. This study has
been done either by using analytical potentials or by using $N$-body
simulations \citep{b6,b1,b2,b3,b7,b4,b8,b5}. The former method is
not self consistent, but it has the advantage of the analytical form
of the potential, allowing the free choice of all the parameters
(like, for example, the strength of the bar and of the spiral
perturbation, or the pattern speed of the bar). The latter method
has the advantage of self consistency, but the time evolving
potential and pattern speed imply a secular evolution of the system
that makes the study harder.

Photometric data have shown that spiral structures are related to
the old stellar disk \citep{b24}. Analytical models and $N$-body
dissipationless simulations have shown that, the chaotic domains are
small in normal spiral galaxies (non barred), because the spiral
perturbations are relatively small. Therefore, in many studies up to
now it was argued that orbits near some stable periodic orbits
(regular motions) can support the spiral pattern  \citep[see for
example][]{b10,b27,b23,b29,b28}.

In \cite{b17} the authors pointed out the role of the so-called `hot
population' \citep{b16}, in supporting the inner parts of the spiral
arms extending beyond the bar. They found chaotic orbits that wander
stochastically, partly inside and partly outside corotation. In
strong barred galaxies the chaotic domains of phase space have been
proved very important \citep[see for example][]{b26}.

The role of the chaotic orbits in the spiral structure of a galaxy
was pursued by \citet{b9}, using self-consistent $N$-body
simulations of barred spiral galaxies, where they found long living
spiral arms composed almost entirely of chaotic orbits.

Most chaotic orbits are connected to the asymptotic manifolds of the
main unstable periodic orbits. \cite{b6} and \cite{b7} have
considered the apocentric positions of the asymptotic orbits and the
coalescence of all the unstable manifolds in a certain range of
energy levels. On the other hand \cite{b2,b3} and \cite{b4,b5} put
the emphasis on the asymptotic manifolds emanating only from the
unstable periodic orbits of families originating at the Lagrangian
points $L_1$ and $L_2$.

In this paper we study a certain snapshot of a 3-D  $N$-body
simulation of a barred spiral galaxy, in order to give a more
concrete and explicit description of the mechanism of construction
of the spiral structure out of sticky chaotic orbits which are close
to asymptotic orbits from various unstable periodic orbits. Since
the $N$-body simulation is a self-consistent process, it gives all
the details of the secular evolution of the galactic system during a
Hubble time, having a potential and a pattern speed that vary with
time. However by using frozen potentials and constant pattern speeds
that correspond to certain snapshots of the evolution we are able to
study the role  of chaotic orbits in supporting a spiral structure
that survives for more than 10 bar rotations, i.e. for at least 1/3
of the Hubble time. The sequence of these snapshots can be
considered as an adiabatic approximation of the real galaxy
evolution.

Chaotic orbits spend a long time close and along the unstable
asymptotic manifolds of the unstable periodic orbits in the phase
space, due to phenomena of stickiness \citep[see][]{b11}. In our
model, there are energy levels where the phase space is not bounded
outwards and chaotic orbits can escape from the system. However, the
escape rate is very small, due to stickiness. The orbits spend most
of their time close to velocity minima on the rotation plane of the
galaxy.

We show that the loci of velocity minima are connected with the
maxima of the density distribution in the configuration space in
every energy level and find the connection of these geometrical loci
with the apocentres and the pericentres of the orbits.

We find the characteristic diagrams of the most important 3-D
periodic families and we reveal the spiral structure of the galaxy
by integrating chaotic orbits that lie close and along the main
unstable periodic orbits.  Finally, we construct the space
configuration by superimposing orbits belonging to different energy
levels. The space distribution of the apocentres and pericentres of
these orbits is very similar to the apocentric and pericentric
manifolds of 2-D asymptotic orbits having initial conditions along
the unstable asymptotic curves of the 2D periodic orbits.

\begin{figure}
\centering
\includegraphics[width=8.5cm]
{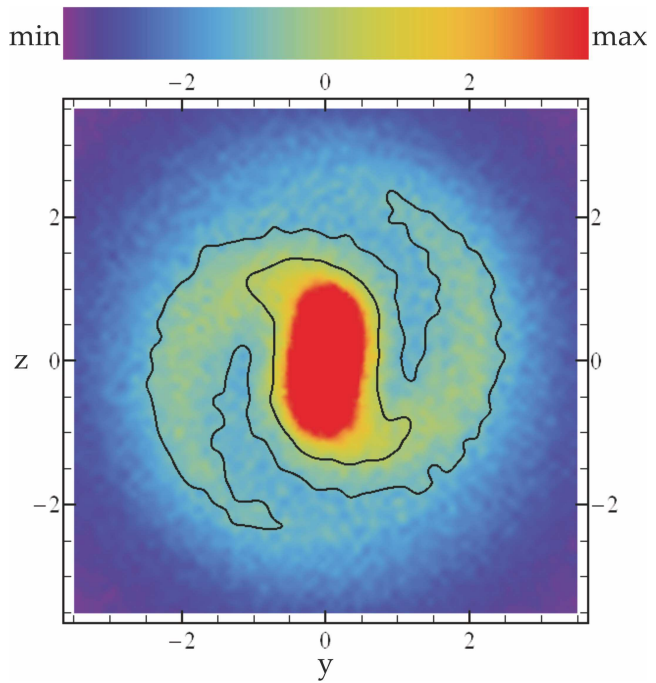} \caption{The density distribution (in the color scale
indicated in the top of the panel) of the projected particles of our
model on the rotation $y-z$ plane made by the superposition of 20
snapshots corresponding to the first 20$T_{hmct}$ of the integration
of the system in the fixed effective potential. Black curves denote
characteristic density contours revealing the spiral structure. Note
that the color bar shown in this figure applies for all the color
scales used in the following figures. } \label{fig01}
\end{figure}

The paper is organized as follows: In section 2 we give a
description of the system providing its main properties. We make a
frequency analysis separately for the chaotic and regular
populations. In section 3 we discuss the role of the apocentres and
the pericentres of the orbits in comparison with the loci of the
minimum velocities on the rotation plane. In section 4 we discuss
the role of the sticky resonant orbits in supporting the spiral
structure of the system. We present examples of specific periodic
orbits that are able to reconstruct the main features of the galaxy.
Moreover, we discuss the role of the apocentric and pericentric
intersections of the 2D asymptotic orbits. In section 5 we give a
description of the way the orbits escape from the system and finally
in section 6 we present our conclusions.

\section[]{Description of the model}

The initial conditions of our $N$-body model, were created by
\cite{b9} where four experiments with different pattern speeds
simulating barred spiral galaxies were produced. In our paper we use
the $QR4$ model of \cite{b9} (the experiment with the greatest value
of the bar's pattern speed).

We use $1.5 \times 10^5$ particles in our simulation. The time unit
is taken equal to the half mass crossing time of the system
(hereafter $T_{hmct}$). A Hubble time corresponds to about $300
T_{hmct}$. The length unit is taken equal to the half mass radius of
the system (hereafter $r_{hm}$). Finally, the plane of rotation is
the $y-z$ plane (intermediate-long axes) and the sense of rotation
is clockwise.

\begin{figure}
\centering
\includegraphics[width=8.5cm]
{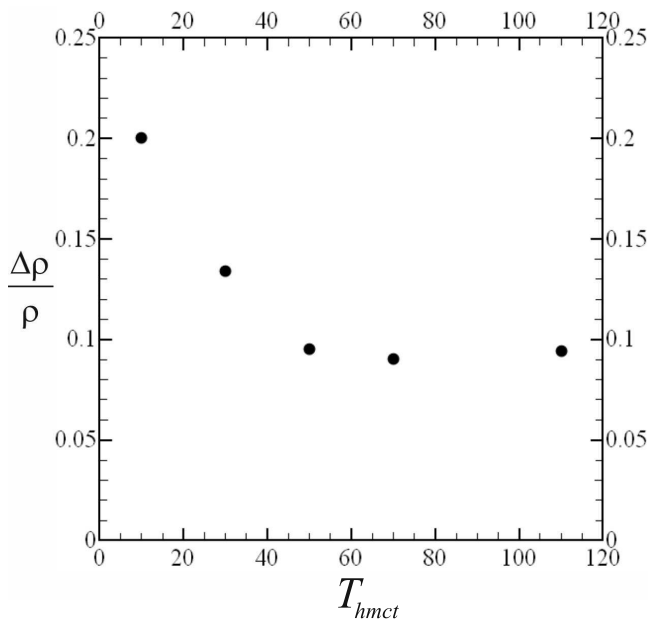} \caption{The time evolution of the amplitude of the
spiral perturbation $\delta \rho/\rho$ at $r\approx 1.7r_{hm}$. We
observe that the spiral structure fades out gradually. }
\label{fig02}
\end{figure}

The Jacobi constant $E_j$ of an orbit is given by the relation
\begin{equation}
E_j=\frac{1}{2}(\dot{x}^2+\dot{y}^2+\dot{z}^2)+V(x,y,z)-\frac{1}{2}
\Omega_p^2 R_{yz}^2 \label{ejacobi}
\end{equation}
where $V(x,y,z)$ is the full 3D `frozen' potential, given by a
Smooth Field Code (SFC code) as an expansion of a bi-orthogonal
basis set, $\dot{x}, \dot{y}, \dot{z}$ are the velocities in the
rotating frame of reference and $\Omega_p$ is the angular velocity
of the bar (or pattern speed) at the studied snapshot. The value
$R_{yz}$ is the distance from the rotation axis
$\left(\sqrt{y^2+z^2}\right)$. The Jacobi constant $E_j$ is called
also ``energy in the rotating frame'' and can be written as
\begin{equation}
E_j=\frac{1}{2}(\dot{x}^2+\dot{y}^2+\dot{z}^2)+V_{eff}(x,y,z)
\label{ejveff}
\end{equation}
where $V_{eff}(x,y,z)=V(x,y,z)-\frac{1}{2}\Omega_p^2 R_{yz}^2$ is
the effective potential.

The energy of the orbit in the inertial frame is related to the
Jacobi constant by the following relation
\begin{equation}
E=E_j+\mathbf{\Omega_p}\cdot\mathbf{L}
\end{equation}
where $\mathbf{L}$ is the angular momentum in the inertial frame of
reference and $\mathbf{\Omega_p}=- \Omega_p\mathbf{i}$, $\mathbf{i}$
being the unit vector along the $x$-axis. Therefore the inertial
energy reads
\begin{equation}
E=E_j-\Omega_p(y P_z-z P_y)
\end{equation}
where $P_y, P_z$ are the momenta in the inertial frame of reference.
Since
\begin{equation}
P_y=\dot{y}+\Omega_p z,~~~~~~P_z=\dot{z}-\Omega_p y
\end{equation}
we have
\begin{equation}
E=\frac{1}{2}(\dot{x}^2+\dot{y}^2+\dot{z}^2)-\Omega_p(\dot{z}y-\dot{y}z)+V(x,y,z)+\frac{1}{2}
\Omega_p^2 R^2_{yz} \label{einert}
\end{equation}

\begin{figure}
\centering
\includegraphics[width=8.5cm]
{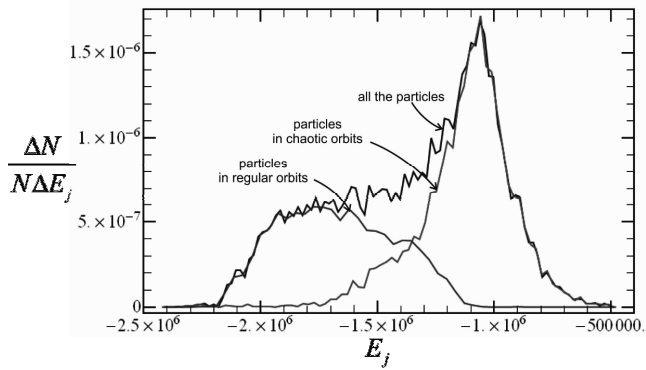} \caption{The distribution of the Jacobi constant values
for all the particles, for the particles moving in regular orbits
and for the particles moving in chaotic orbits. The maximum of the
distribution corresponds to chaotic particles around corotation. }
\label{fig03}
\end{figure}

Our study is made in a distinct snapshot of the $N$-body simulation
that corresponds to 55$T_{hmct}$. At this snapshot the spiral
structure of the galaxy is clearly visible and moreover it survives
for several rotations of the bar. In order to study the mechanism
that creates this spiral structure which lasts for such a long time
we fix the potential and the pattern speed and we study the role of
chaotic orbits in this fixed potential. The corresponding value of
the pattern speed is $\Omega_p\approx2\pi/8T_{hmct}$ and is
calculated in \cite{b9} (see Fig.~4a of that paper). In real units
this value corresponds to $\approx25km~ sec^{-1}~ kpc^{-1}$. This
value is in the range of the values observed in late type barred
spiral galaxies \citep[see for example][]{b25}.

By integrating all the orbits in this fixed effective potential we
reveal the longevity of the spiral structure for at least
$100T_{hmct}$, while after much longer times this structure
disappears.

In Fig.~1 we plot, on the rotation plane $y-z$, the superposition of
20 snapshots corresponding to the first $20T_{hmct}$ of the
integration time of the system in the fixed effective potential.
Such a superposition consists of more than $2.5\times 10^6$
particles and allows us to reduce significantly the noise that
appears when the number of particles is small. Since the density is
now clearly defined, we can safely quantify the spiral perturbation.
In Fig.~2 we show the time evolution of the amplitude of the spiral
perturbation in our model. More precisely, we plot the density
excess ($\delta\rho/\rho$) caused by the spiral structure in a
specific radius that corresponds to $\approx 1.7r_{hm}$ as a
function of time. After a time period of $100T_{hmct}$, during which
the bar has completed $\approx 13$ rotations, the spiral
perturbation is still marginally detectable. This time corresponds
to 1/3 of the Hubble time.

Figure 3 gives the distribution of the Jacobi constants
(Eq.~\ref{ejveff}) for all the particles and separately for chaotic
and for regular orbits. The distinction between the two populations
is done by the method introduced by \cite{b9}. The percentage of
chaotic orbits is found to be $\approx 60\%$. The maximum of the
distribution of the chaotic orbits is found between the values
$E_j(L_1)\approx-1120000$ and $E_j(L_4)\approx-1080000$
corresponding to the Jacobi constant values of the Lagrangian points
$L_{1,2}$ and $L_{4,5}$ respectively. From Fig.~3 it is obvious that
the regular motions are restricted inside corotation and therefore
they support the shape of the bar, while chaotic orbits are
responsible for the spiral structure outside corotation.

\begin{figure}
\centering
\includegraphics[width=8.5cm]{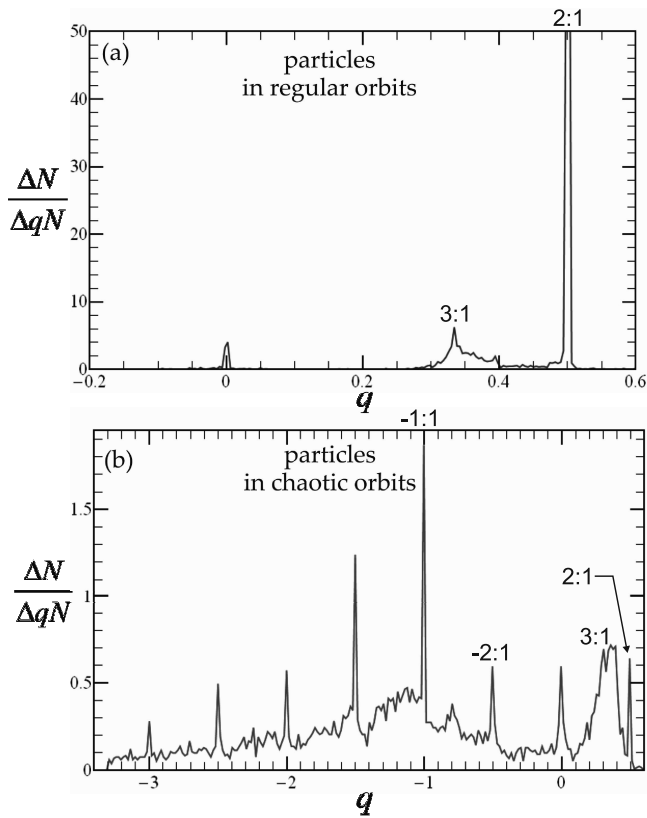}
\caption{ The frequency analysis of \textbf{(a)} the regular
component and \textbf{(b)} the chaotic component of the system. The
frequency ratio $q$ is given by Eq. (7). We see that the vast
majority of regular orbits lie inside corotation $(q>0)$ (especially
around 2:1 resonance) while the chaotic orbits lie both inside
$(q>0)$ and outside corotation $(q<0)$. The various spikes shown in
\textbf{(b)} correspond to chaotic orbits sticking close to various
resonances. } \label{fig04}
\end{figure}

Figure 4 presents the frequency analysis of the regular component
(Fig.~4a) and of the chaotic component (Fig.~4b) of the system. We
are only interested in revealing the ``disc'' resonances on the
rotation plane and therefore the frequency analysis is made in the
2D projection of the orbits on the $y-z$ plane. However, there exist
also vertical resonances that influence mainly the edge-on profiles
of our systems \citep[see][]{b22}.

\begin{figure}
\centering
\includegraphics[width=8.5cm]
{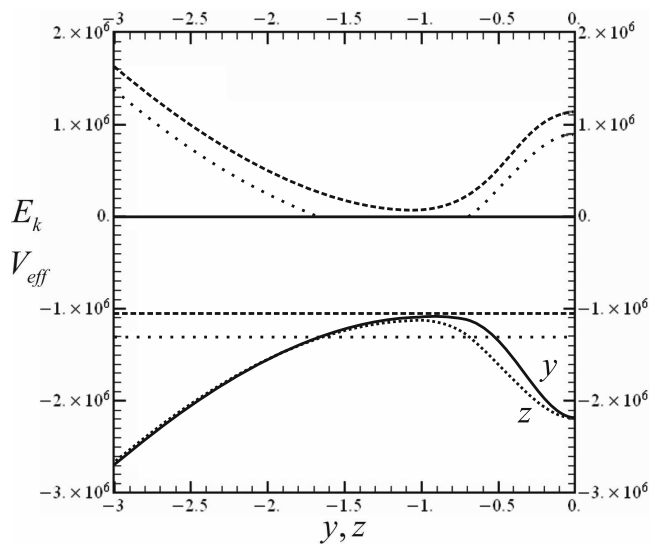} \caption{The effective potential $V_{eff}$ along the
$y$-axis (black solid line) and along the $z$-axis (black dotted
line). The two curves of kinetic energy $E_{k}$ (dotted and dashed
curves) correspond to two different energy levels
$E_j=-1300000<E_j(L_1)$ (dotted line) and $E_j=-1050000>E_j(L_1)$
(dashed line). In the former case the minimum velocity corresponds
to apocentres for the motion inside corotation and to pericentres
for the motion lying outside corotation. In the latter case the
minimum velocity corresponds to the position of the maximum of
$V_{eff}$.} \label{fig05}
\end{figure}

The ``frequency ratio'' $q$ is a parameter given by the relation
\begin{equation}
q=\frac{\Omega-\Omega_p}{\kappa}
\end{equation}
where $\Omega$ is the angular velocity of a particle in the inertial
frame of reference and $\kappa$ is the epicyclic (radial) frequency
\citep{b31}. Positive $q$ values correspond to resonances inside
corotation while negative $q$ values correspond to resonances
outside corotation, which are related only to chaotic orbits. In
Fig.~4a we see that the main resonance of the regular orbits (inside
corotation) is the 2:1 resonance (or inner Lindblad resonance),
while there is also a small number of particles around the 3:1
resonance and an even smaller number around corotation $(q=0)$. The
vast majority of these orbits support the bar of the system. Figure
4b shows that a rather small percentage of chaotic orbits inside
corotation lie near important resonances (e.g. 2:1, 3:1 and 4:1).
These represent sticky chaotic orbits mainly supporting the outer
layers of the bar \citep[see][]{b8} and the innermost parts of the
spiral structure. The rest of the chaotic orbits move around
corotation and outside it and some of them stick around specific
resonances ($q$=0, -0.5, -1, -1.5, -2, -2.5, -3). The $q=0$
population corresponds to chaotic orbits located around corotation
and  close to the $PL_1$, $PL_2$, $PL_4$ and $PL_5$ families
\citep[nomenclature after][]{b6}. Below we will show that the
chaotic orbits sticking along the asymptotic manifolds originating
from these unstable periodic orbits are responsible for the spiral
structure of the system.

\section{The role of the apocentres and pericentres}

In this section we discuss the role of apocentres or pericentres in
revealing the morphological features of the system and we relate
their geometrical loci with the locus of the velocity minima on the
rotation plane.

The distribution of the radial velocities $\dot{r}$ of the $N$-body
orbits has a preferable concentration around the value zero,
indicating that the apocentres and pericentres should play an
important role for the system.

The geometrical loci of the minima of the velocities $v_{yz}$ on the
rotation plane should correspond to the maxima of the density
distribution on the configuration space, because particles spend
most of their time there. It is of interest to investigate how these
loci are related to the apocentres or the pericentres of the orbits.

\begin{figure*}
\centering
\includegraphics[width=\textwidth]
{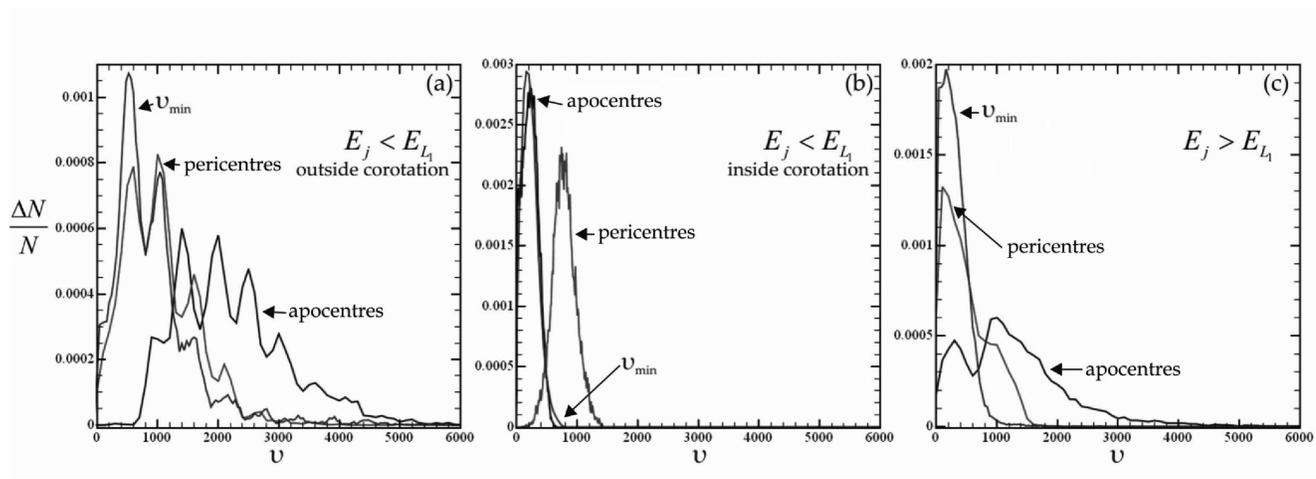} \caption{The distribution of the velocities of the
apocentres, of the pericentres and of the velocity minima for
\textbf{(a)} $E_j<E_j(L_1)$ outside corotation \textbf{(b)}
$E_j<E_j(L_1)$ inside corotation and \textbf{(c)} for $E_j>E_j(L_1)$
where the areas inside and outside corotation can communicate.}
\label{fig06}
\end{figure*}

\begin{figure*}
\centering
\includegraphics[width=\textwidth]
{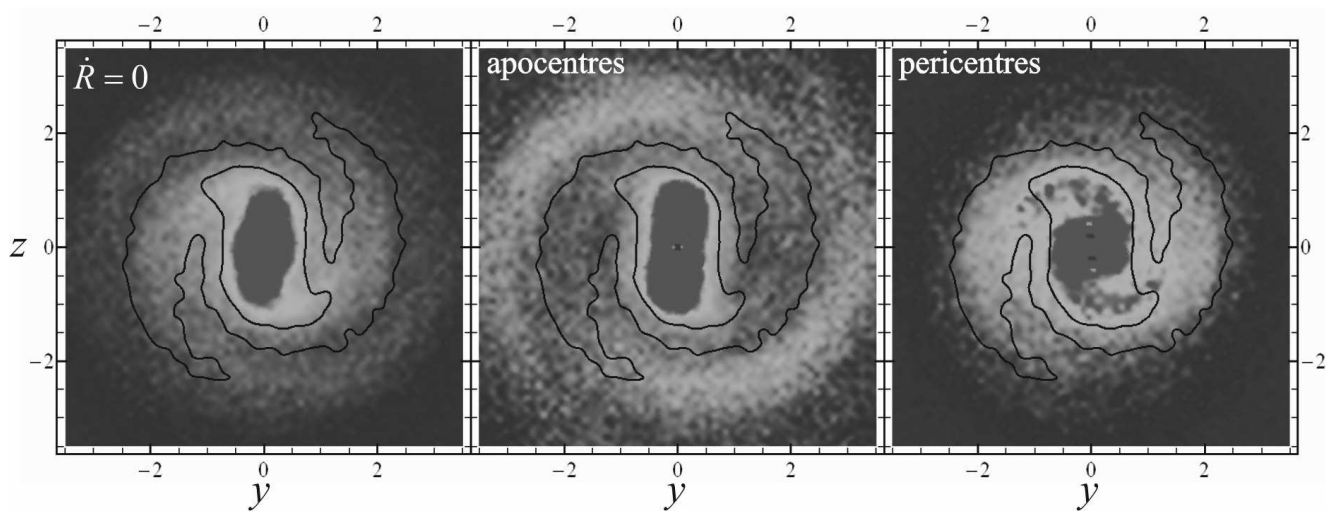} \caption{\textbf{(a)} The density distribution of the
particles of Fig.~1 when integrated until their (cylindrical) radial
velocities become zero ($\dot{R}=0$) for the first time, i.e. at the
apocentres or the pericentres of their orbits. The density contours
corresponding to the initial conditions of Fig.~1 are superimposed
(black curves). \textbf{(b)} The density distribution of only the
apocentres. The comparison with the density contours shows that the
apocentres support the bar but not the extensive spiral arms.
\textbf{(c)} The density distribution of only the pericentres. The
comparison with the density contours shows that the pericentres
support the spiral structure but not the bar.} \label{fig07}
\end{figure*}

In Fig.~5 we plot the effective potential of a 2D approximation of
the system ($x=0$) along the $z$-axis
($V_{eff}(0,0,z)=V(0,0,z)-\frac{1}{2} \Omega_p^2 z^2$, black dotted
curve) and along the y-axis ($V_{eff}(0,y,0)=V(0,y,0)-\frac{1}{2}
\Omega_p^2 y^2$, black solid curve). The maximum of the black dotted
curve corresponds to the Jacobi constant $E_j(L_1)$ of the
Lagrangian points $L_{1,2}$, while the maximum of the black solid
curve corresponds to the Jacobi constant $E_j(L_4)$ of the
Lagrangian points $L_{4,5}$. For any Jacobi constant value below
$E_j(L_1)$, for example for $E_j=-1300000$ (dotted horizontal
straight line), there is a forbidden region for the orbits (region
between the two points of intersection with the effective potential
curve). In this case, the phase space, as well as the configuration
space, consists of two regions that do not communicate with each
other (the one inside and the other outside corotation). By plotting
the corresponding curve of the kinetic energy (dotted curve) and
because of (Eq.~\ref{ejveff}), we see that for the region inside
(outside) corotation the velocity minima correspond to the
apocentres (pericentres) of the orbits. However, for $E_j$ values
above $E_j(L_4)$, for example for $E_j=-1050000$ (dashed horizontal
straight line), the region inside corotation can communicate with
the region outside corotation and the minimum velocity $v_{min}$
(minimum of the dashed curve) is found at the position of the
maximum of the effective potential (i.e. the minimum difference of
$E_j-V_{eff}$). In this case, the geometrical locus of the velocity
minima on the configuration space is not directly related to the
apocentres or the pericentres of the orbits.

\begin{figure}
\centering
\includegraphics[width=8.5cm]
{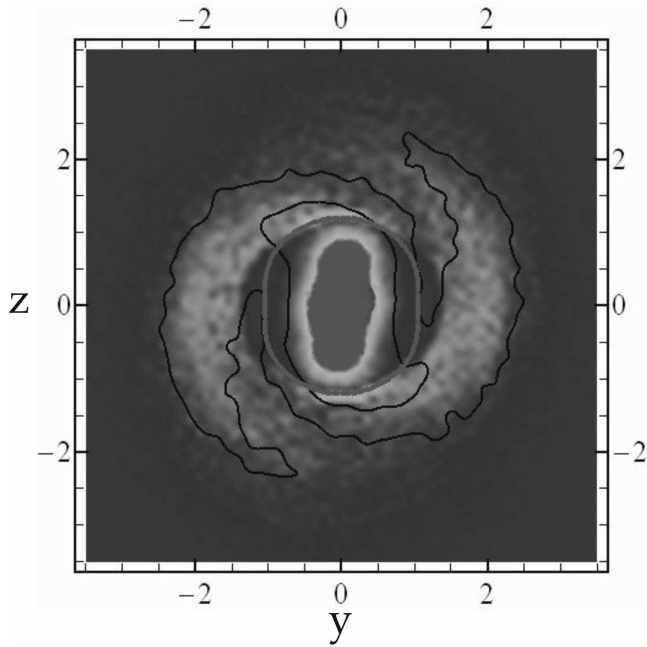} \caption{The density distribution of the particles of
Fig.~1 (with $E_j<E_j(L_4)$) at the positions corresponding to their
local plane velocity minima ($v_{yz}=v_{min}$). We observe that this
distribution fits well the morphology of the galaxy (black contour
lines). In this figure we have omitted the particles with
$E_j>E_j(L_4)$ which are gathered along the geometrical locus of the
maximum of the effective potential (gray elliptical curve).}
\label{fig08}
\end{figure}

Figure 6 presents the distribution of the velocities of the
pericentres of the apocentres and of the velocity minima for three
cases: \textbf{(a)} for orbits trapped outside corotation with
$E_j<E_j(L_1)$, (Fig.~6a), \textbf{(b)} for orbits inside corotation
with $E_j<E_j(L_1)$, (Fig.~6b) and \textbf{(c)} for orbits that can
explore the whole permissible phase space inside and outside
corotation with $E_j>E_j(L_1)$, (Fig.~6c).

In Fig.~6a we observe that the pericentres of the orbits outside
corotation have smaller velocities than the apocentres. Moreover the
distribution of the velocity minima gives good agreement with the
distribution of the velocities of the pericentres. At the same time,
the distribution of the velocities of the apocentres almost coincide
with the distribution of the velocity maxima (not plotted in the
panel). Therefore, the particles outside corotation spend most of
their time near the pericentres of their orbits which is consistent
with the description of Fig.~5. On the other hand the opposite is
true for particles inside corotation (Fig.~6b). The particles inside
corotation have smaller velocities at their apocentres than at their
pericentres. Moreover, the apocentric velocity distribution almost
coincides with the distribution of the velocity minima. Here again
the distribution of the velocity maxima (not plotted in the panel)
almost coincides with the distribution of the velocities of the
pericentres.

For the orbits having Jacobi constant above $E_j(L_1)$ (Fig.~6c)
even though there is not a clear breakoff of the distributions, the
pericentres have statistically smaller velocities than the
apocentres. Both pericentres and apocentres present a peak around
the same value of velocity, which also coincides with the peak of
the velocity minima distribution. The distribution of the
apocentres, however, has a second important peak around a somewhat
greater value of velocity.

In Fig.~7a we plot, in color scale, the density distribution of the
particles of Fig.~1 taking each of them at a position corresponding
to $\dot{r}=0$, i.e. at the apocentre or the pericentre of its
orbit. The black curves correspond to the density contours of the
space distribution shown in Fig.~1. Figure 7b presents, the density
distribution of only the apocentres of the orbits. The corresponding
density maxima inside corotation support the bar (since they
correspond to loci of minimum velocities, see Fig.~6b) as well as a
small extension of the bar to the inner parts of the spiral
structure. However, the density maxima outside corotation, do not
correspond to the density contours of the real spiral arms (black
curves) but they extend far beyond. Figure 7c presents the density
distribution of only the pericentres of the orbits. We see that
these density maxima support a spiral structure very close to the
real one, but a little more tight, and they do not support the bar.

Thus, we conclude that, in general, the apocentres of the orbits
support the shape of the bar and the origins of the spirals near
corotation, while the pericentres of the orbits better support the
spiral structure near and outside corotation.

Figure 8 presents, in gray scale, the density distribution of the
particles of Fig.~1 at positions corresponding to local minimum
values of their rotation plane velocities ($v_{yz}=v_{min}$). The
comparison with the density contours (black curves) of the initial
conditions shows that the loci of the velocity minima reveal
satisfactorily the features of the galaxy, i.e. the bar and the
spiral arms. In this figure, however we have removed the overdensity
of the particles along the theoretical locus of $v_{min}$ (gray
elliptical curve) that represents the maximum of the effective
potential ($V_{eff}$) for Jacobi constants above $E_j(L_4)$, because
it does not correspond to the distribution of the real particles.

\begin{figure*}
\centering
\includegraphics[width=15.cm]
{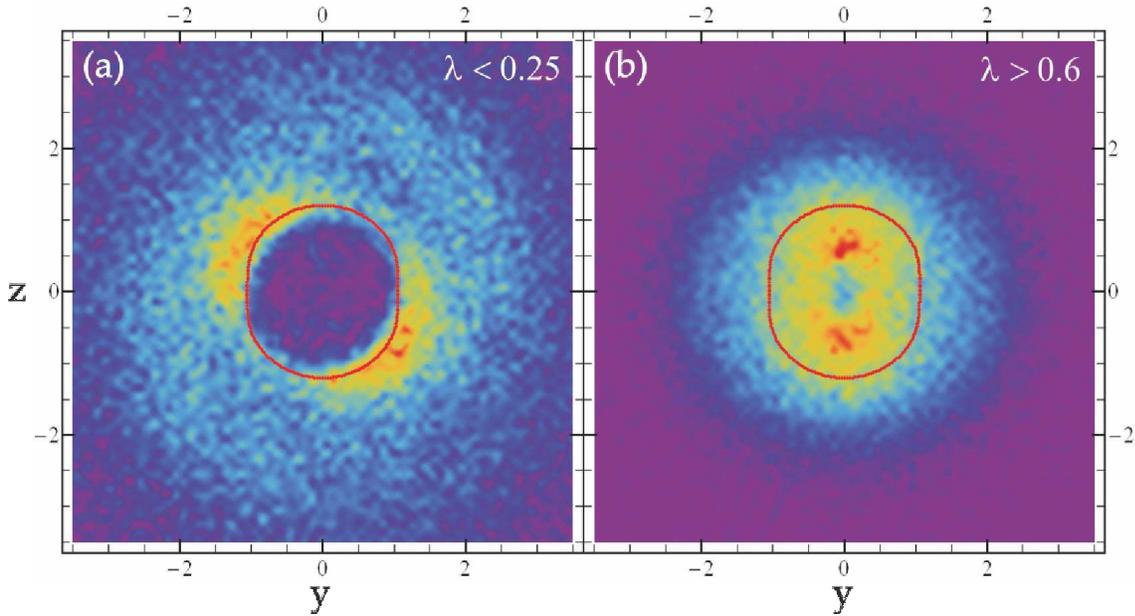} \caption{\textbf{(a)} The density distribution of the
particles having $E_j>-977000>E_j(L_4)$ and ratio $\lambda=\frac
{<v_{\min}>}{<v_{yz}>}<0.25$ together with the locus corresponding
to the effective potential maxima $V(eff)_{max}$ (red curve). The
density maxima in this case are located near the curve of
$V(eff)_{max}$ but along particular directions. \textbf{(b)} The
same as in \textbf{(a)} but for particles having
$\lambda=\frac{<v_{\min}>}{<v_{yz}>}>0.6$. The density maxima are
located inside the curve of $V(eff)_{max}$ (see text for details). }
\label{fig09}
\end{figure*}

\begin{figure*}
\centering
\includegraphics[width=15.0cm]
{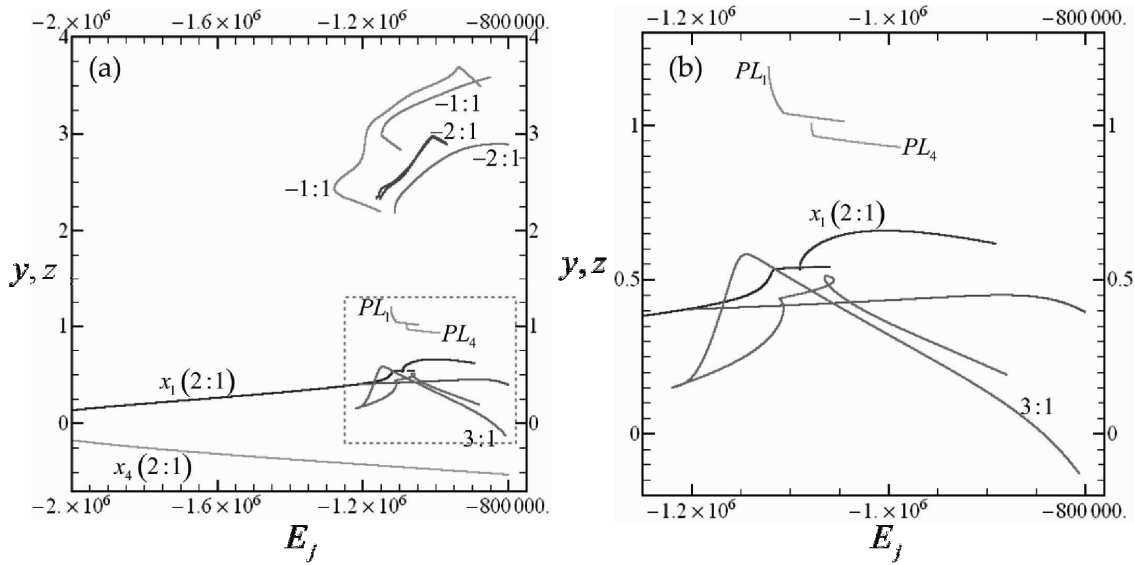} \caption{\textbf{(a)} The characteristics of the main 3D
periodic orbits of the system. \textbf{(b)} A focusing in the
inserted frame of \textbf{(a)}.} \label{fig10}
\end{figure*}

Here we note that a density maximum in some area can be produced
either by frequent (but not too short-lasting) transits or by
long-lasting (but not too rare) transits of the particles through
this area. For Jacobi constant values higher than $E_j(L_4)$ the
locus of the local velocity minima (red curve in Fig.~8) corresponds
to velocity values that can be comparable to the mean velocity of
the whole orbits. In Fig.~9 we have plotted the density distribution
of the particles with Jacobi constant values $E_j>-977000>E_j(L_4)$
corresponding to low and high values of the ratio
$\lambda=<v_{\min}>/<v_{yz}>$ as indicated on the two panels of
Fig.~9. We observe that the particles with low $\lambda$ values
(Fig.~9a) form a ring-like density maximum area lying close to the
locus of the maximum $V_{eff}$ (red curve). However the distribution
is not uniform along the red curve, but it follows the distribution
of the inner parts of the spiral arms. The maximum density area of
the particles with high $\lambda$ values (Fig.~9b) is located inside
the red curve corresponding to the velocity minima. In this case the
density maxima are due to the more frequent transits of the orbits
from this area.

\section{Sticky Chaotic Orbits}

In Figure 4 the frequency ratio $q$ of the chaotic population is
plotted showing preferable concentrations around specific resonant
periodic orbits. In order to study the behavior of sticky resonant
orbits we find the characteristics of the most important periodic
orbits in the 3D case. Figure 10 presents the characteristics of the
main 3D periodic orbits, i.e. the $y$ ($z$ for the $PL_4$ family)
component as a function of the Jacobi constant. We must point out
that the periodic orbits $PL_1$ and $PL_2$ exist only above the
value $E_j(L_1)=-1120000$, while the periodic orbits $PL_4$ and
$PL_5$ exist above the value $E_j(L_4)=-1080000$. We will show below
that all the important resonances, presented in the characteristic
diagram (Fig.~10) and in the frequency analysis of the chaotic
orbits (Fig.~4b), contribute to the formation of all the
morphological features appearing in each Jacobi constant value.

\begin{figure*}
\centering
\includegraphics[width=\textwidth]
{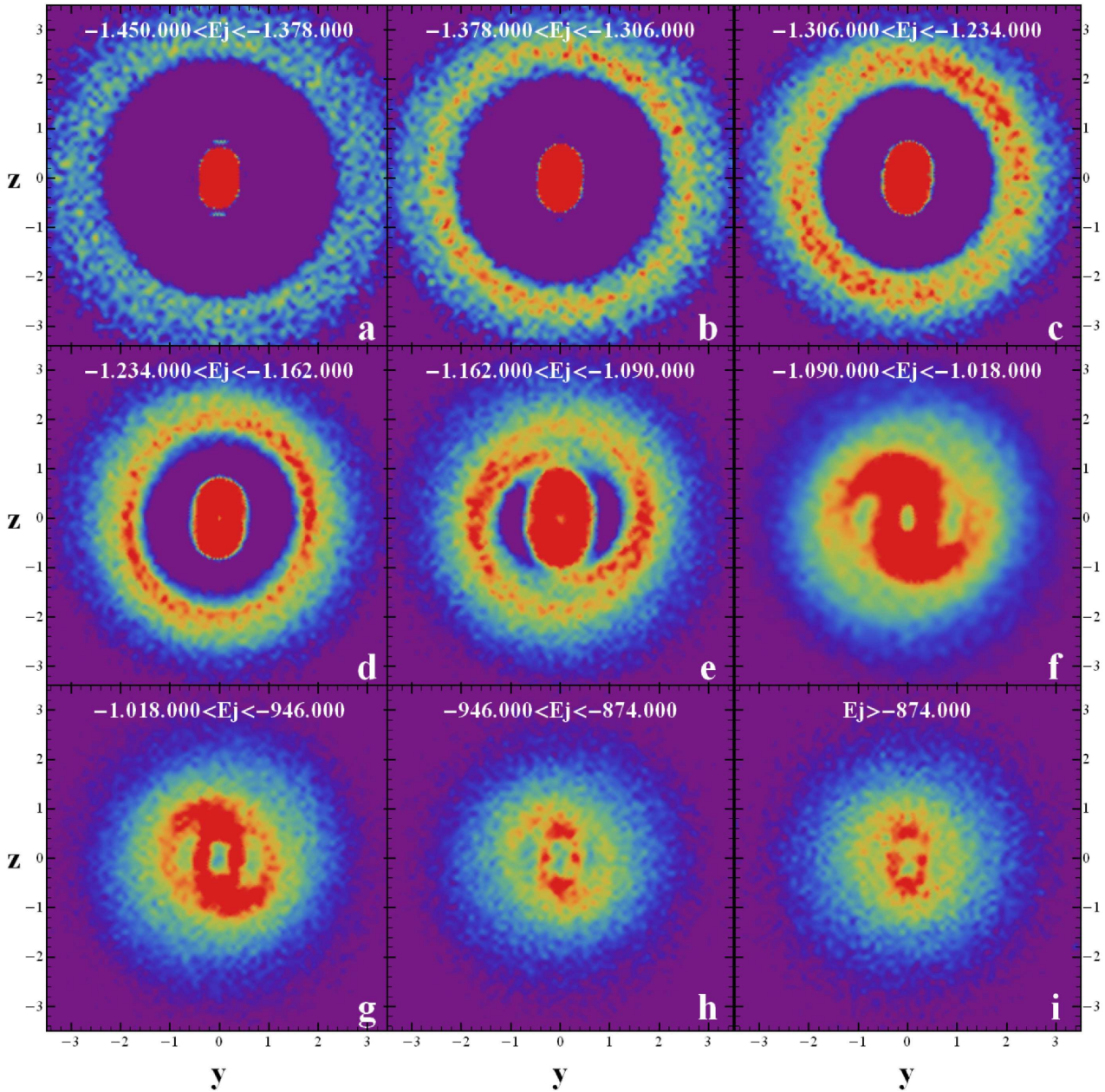} \caption{The density distribution on the $y-z$ plane of
the particles belonging to \textbf{nine} different Jacobi constant
bins. The corresponding values are marked on the top of each panel.
The top left panel corresponds to the smallest values of Jacobi
constants $E_j$ (close to the potential well) while the bottom right
panel corresponds to the highest values of $E_j$'s. In this figure
we reveal which structures are supported by the particles
corresponding to the different Jacobi constant values.}
\label{fig11}
\end{figure*}

Figure 11 presents, in color scale, the density distribution of
particles on the rotation plane for nine different Jacobi constant
levels. The top left panel corresponds to the lowest values of
Jacobi constants $E_j$ (close to the potential well) while the
bottom right panel corresponds to the highest values of $E_j$'s. The
corresponding values of the Jacobi constant are marked on the top of
each panel. Note that outside corotation (in energy levels with
$E_j>E_j(L_1)$) there exist practically only chaotic orbits (see
Fig.~3). A general trend is that the space distribution of particles
having lower values of Jacobi constants supports mostly the outer
parts of the spiral arms as well as the shape of the bar, while
particles having greater values of Jacobi constants support the
inner parts of the spiral arms. This is obvious in Figs. 11 b-h,
where we see a gradual shift of the maximum of the density
distribution of particles outside corotation from the outer parts of
the spiral arms to the inner parts and the innermost extensions of
the bar (Figs. 11f,g). On the other hand there are levels of the
Jacobi constant that do not support at all the spiral structure.
This happens for the lowest values of the Jacobi constant near the
potential well (Fig. 11a), where the distribution of particles
outside corotation is quite uniform, and for the highest values of
the Jacobi constant (Fig. 11i), where only chaotic orbits exist,
which do not support the spiral structure but rather the outer parts
of the bar.

Below we give some examples of the effect of stickiness
\citep[][]{b11} of chaotic orbits around the unstable asymptotic
manifolds of the unstable periodic orbits, in the phase space. We
show that this stickiness supports the formation of the spiral
structure, while the (slow) diffusion along the paths of the
manifolds is responsible for the fadeout of this structure.

In the next subsections we give four examples of 3D chaotic orbits
lying near and along some main unstable periodic orbits, having
initial conditions inside corotation, near corotation and outside
corotation.

\begin{figure*}
\centering
\includegraphics[width=\textwidth]
{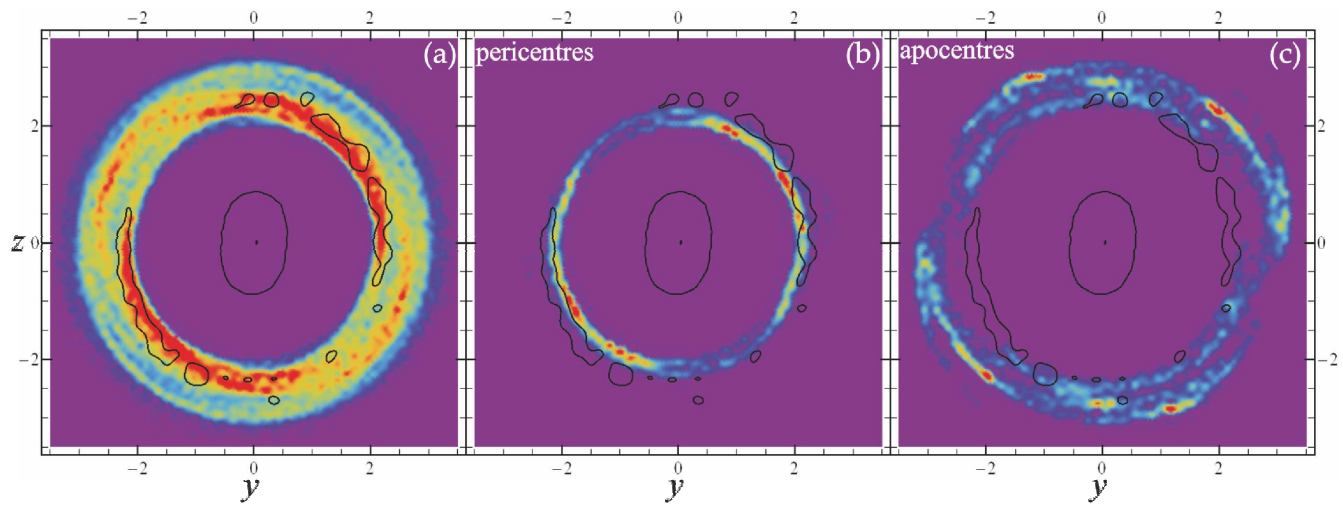} \caption{\textbf{(a)} The density distribution on the
$y-z$ plane of 20000 3D orbits integrated for several $T_{hmct}$
having initial conditions close to and along the unstable periodic
orbit $-1:1$ and Jacobi constant $E_j=-1252560$. Black curves
indicate the high density contours of real particles corresponding
to the same energy level. We note the coincidence of these density
contours with the density maxima of the integrated orbits.
\textbf{(b)} the density distribution of the pericentres
superimposed with the high density contours of the corresponding
energy level. \textbf{(c)} the same as in \textbf{(b)} but for the
apocentres. We observe that the pericentres (apocentres) fit (do not
fit) the density maxima of the real particles.} \label{fig12}
\end{figure*}

\subsection{The -1:1 family}

Figure 12 gives an example of the density distribution of orbits
around an unstable periodic orbit of the $-1:1$ family with
$E_j=-1252560$ (within the range of the top-right panel of Fig.~11)
where the motion is limited outside corotation. We have taken 20000
initial conditions all along this 3D unstable periodic orbit (and
its symmetric with respect to the origin), with small deviations
from it, on the 6-dimensional space (${x,y,z,v_x,v_y,v_z}$). Then we
integrated all these orbits for a time corresponding to many
rotations of the bar. In Fig.~12a we plot the density distribution
of these orbits on the rotation plane. In this figure we have
considered a time interval within 5-6 bar rotations. In the same
figure we have superimposed the high density contours of the
corresponding Jacobi constant (black curves) and we see that the
orbits close to the $-1:1$ periodic orbit reproduce well the density
distribution giving the outer parts of the spiral structure. In
Fig.~12b(12c) we plot the density distribution of the pericentres
(apocentres) of the orbits of Fig.~12a together with the high
density contours (black curves) corresponding to this $E_j$ value.
On the one hand the pericentric density maxima seem to follow a part
of the spiral structure but at a slightly smaller radius. On the
other hand the apocentric distribution does not fit well the real
density maxima. Finally we remark that the loci of the velocity
minima practically coincide with the ones of the pericentres.

\subsection{The 3:1 family}

Figure 13 gives a similar information as Fig.~12, but for an
unstable periodic orbit of the 3:1 family, which is located inside
corotation for $E_j=-1090000$ (within the range of the middle-right
panel of Fig.~11). This Jacobi constant value, is close to
$E_j(L_1)$, and the areas inside and outside corotation can
communicate. We considered again 20000 initial conditions all along
this 3D $(3:1)$ unstable periodic orbit (and its symmetric with
respect to the origin), with small deviations from it, in the
6-dimensional space (${x,y,z,v_x,v_y,v_z}$). These orbits follow, in
the phase space, the unstable directions of the asymptotic manifolds
originating from the periodic orbit. For a short time (the shortness
depends on how large is the initial divergence from the periodic
orbit) these orbits stay very close to the periodic orbit, but later
they start deviating considerably from it. In Fig.~13a we see the
density distribution (in color scale) of these orbits after within a
time interval of 3-4 bar rotations. The black curves indicate the
high density contours of the real particles for the same Jacobi
constant value. It is evident that these two distributions fit one
another very well. A similar behavior has been found by \cite{b18}
for chaotic orbits near the 4:1 resonance. This type of orbits was
considered responsible for the inner parts of the spiral arms in an
analytical model representing the spiral galaxy NGC 4314.

In Fig.~13b we have plotted the positions of all the orbits for the
time interval of Fig.~13a. The color of each point shows the
corresponding velocity value on the rotation plane. The color scale
is the same as for the densities (see Fig.~1) which means that blue
color corresponds to the minima and red color to the maxima of the
velocity values. We observe that the density maxima are well
correlated with the areas of small velocity values (dark to light
blue colors in Fig.~13b). Figures 13c,d are similar to Figs.~12b,c.
We observe that the pericentres of the orbits (Fig.~13c) form a
slightly more closed spiral structure than that of the real
particles (black density contours). On the other hand the apocentres
of the orbits (Fig.~13d) fit well only the density contours of the
bar and the innermost parts of the spiral structure that originate
from the end of the bar. In Fig.~13e we plot the density
distribution of the loci of the local velocity minima. We notice
that this distribution matches better the distribution of the
apocentres in what concerns the bar and the distribution of the
pericentres in what concerns the spiral structure.

\subsection{The 2:1 ($x_1$) family}
The 2:1 \citep[or $x_1$ after the nomenclature of][]{b13} family is
generally considered responsible for the formation and the
robustness of the bar. This is due to its morphology, together with
the fact that this family is usually stable for low values of the
Jacobi constant. According to this point of view the bar should end
not beyond the point where the $x_1$ family turns from stable to
unstable. Here we  point out the usefulness of the unstable part of
this family which supports the spiral structure. Figure 14a,b is
similar to Fig.~13a,b but for the $x_1$ family and for the same
Jacobi constant value. We observe that the density of the orbits
starting close to the $x_1$ periodic orbit is similar to the density
of the orbits with initial conditions near the 3:1 periodic orbit.
Their density maxima support the morphological structures (bar and
spiral) of the real particles for the same Jacobi constant value
(Fig.~14a). These maxima lie near the areas corresponding to low
values of plane velocities $v_{yz}$ (Fig.~14b).

\begin{figure*}
\centering
\includegraphics[width=\textwidth]
{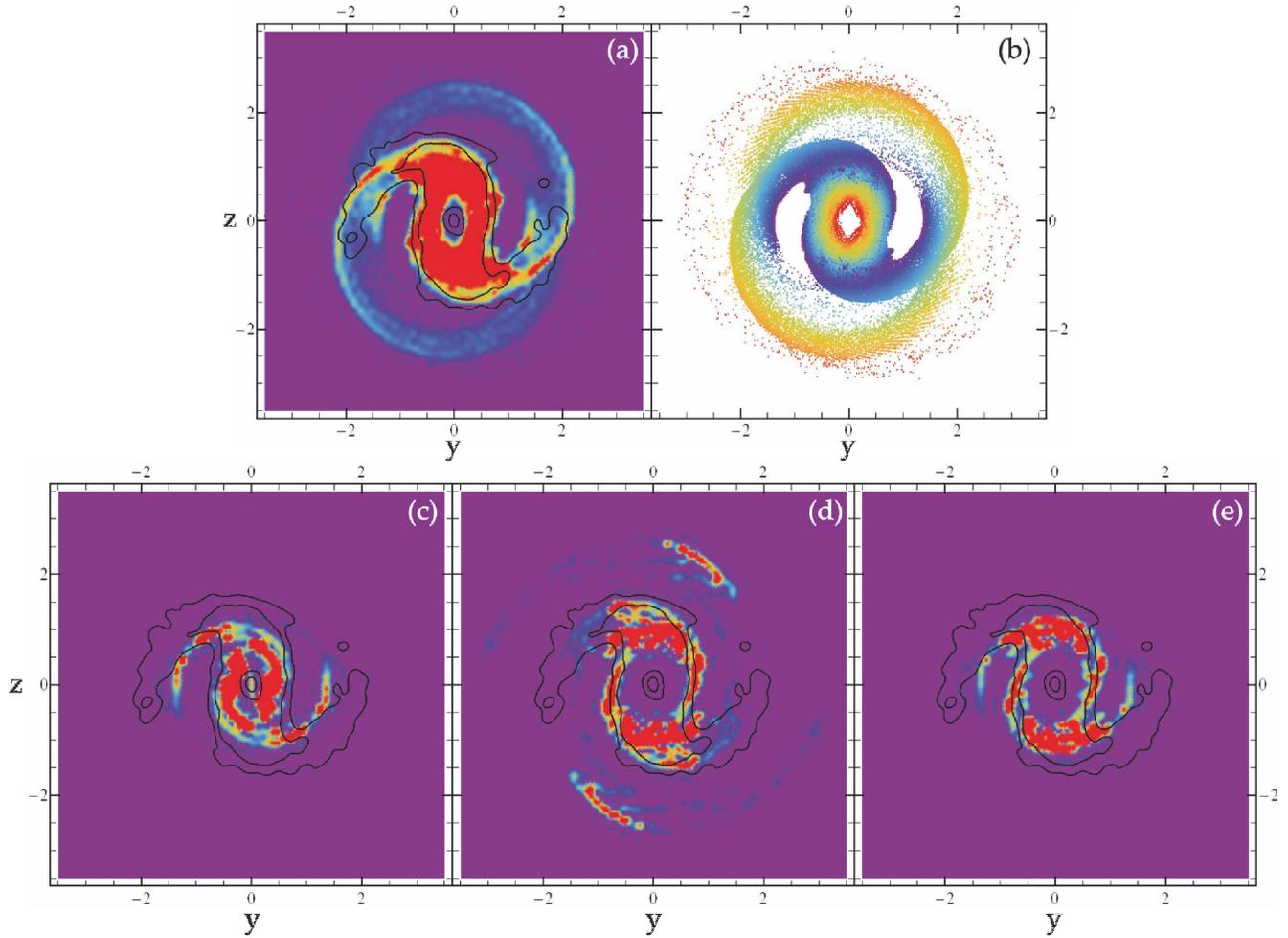} \caption{\textbf{(a)} The density distribution on the
$y-z$ plane of 20000 3D orbits having initial conditions close to
and along the unstable periodic orbit 3:1 (and its symmetric with
respect to the origin) at $E_j=-1090000$. The plotted distribution
corresponds to a time interval within 3-4 bar rotations. Black
curves denote density contours of the real particles corresponding
to the same Jacobi constant value. \textbf{(b)} The positions of all
the orbits at the snapshot shown in \textbf{(a)}. The color of each
point indicates the corresponding velocity value on the rotation
plane. The color scale is the same as for the densities (see
Fig.~1). \textbf{(c)} The density distribution of the pericentres
superimposed with the density contours of the corresponding Jacobi
constant value. \textbf{(d)} The same as in \textbf{(c)} but for the
apocentres. \textbf{(e)} The same as in \textbf{(c)} but for the
local velocity minima $v_{min}$. } \label{fig13}
\end{figure*}

\subsection{The $PL_{1,2}$ families}

Figure 15 is similar to Fig.~14 but for the unstable periodic orbits
$PL_{1,2}$ which are located near corotation and for the same Jacobi
constant value. In this case we establish the same behavior (as for
the cases 3:1 and $x_1$) for the orbits initiating close to the
symmetric periodic orbits $PL_{1,2}$. The orbits starting near the
periodic orbits 3:1, 2:1, $PL_{1,2}$ follow the unstable directions
of the asymptotic manifolds originating from these periodic orbits.
For the same Jacobi constant value these manifolds cannot intersect
each other, and this means that these manifolds are parallel in the
phase space. Thus, the orbits along the unstable directions of the
various manifolds follow parallel paths and present a similar
behavior in the configuration space. The major difference between
the different sets of orbits starting near the different unstable
periodic orbits (3:1, 2:1, $PL_{1,2}$) is the diffusion rate (see
section 5), which increases towards higher resonance orders
$(2:1\rightarrow 3:1\rightarrow \infty:1\equiv PL_{1,2})$.

\begin{figure*}
\centering
\includegraphics[width=12.0cm]
{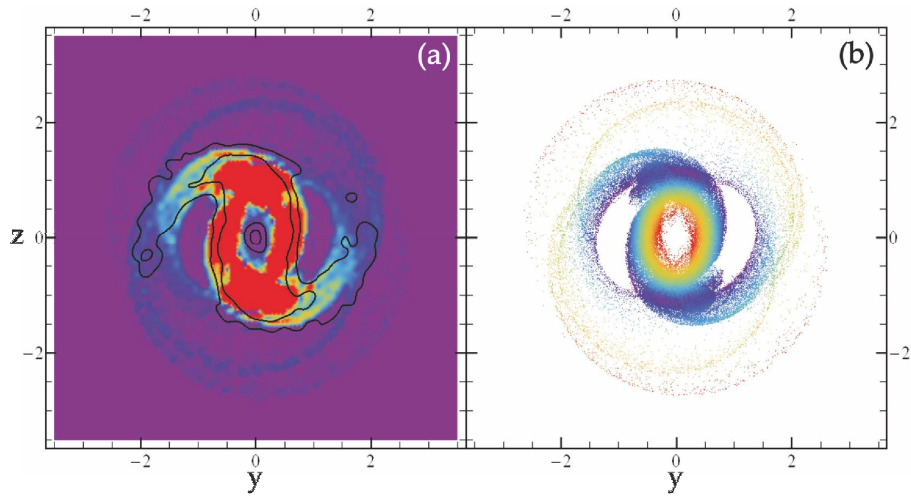} \caption{Similar to Figs.~13a,b but for the unstable
periodic orbit 2:1 $(x_1)$ and for the same Jacobi constant value.}
\label{fig14}
\end{figure*}

\begin{figure*}
\centering
\includegraphics[width=12.0cm]
{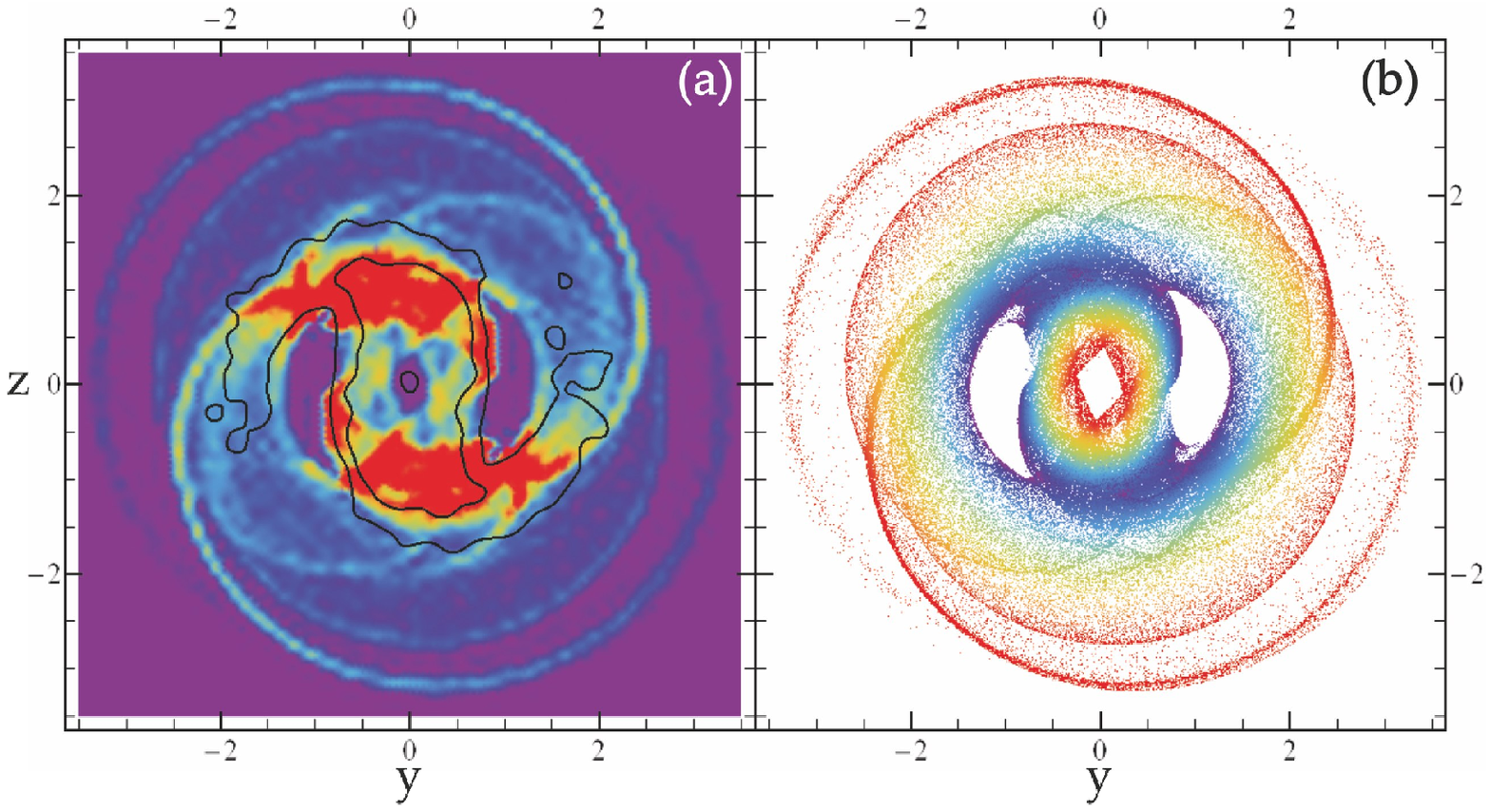} \caption{Similar to Figs.~13a,b but for the unstable
periodic orbits $PL_{1,2}$ and for the same Jacobi constant value.}
\label{fig15}
\end{figure*}

\subsection{Other families}
All the other families of periodic orbits have similar behavior to
those studied in Figs.12-15. Thus, the behavior of the orbits with
$-1300000\lesssim E_j<E_{L_1}$ lying outside corotation (e.g.
$-2:1$) resembles that of the $-1:1$ family (Fig.~12) supporting the
outer parts of the spiral structure. Families lying inside
corotation with $E_j>E_{L_1}$ (e.g. 4:1, $PL_{4,5}$) behave
similarly to the families 2:1, 3:1, $PL_{1,2}$ (Figs.~13-15) and
support the bar and the inner parts of the spiral arms connected to
the bar. Note that there are chaotic orbits associated with the 2:1,
3:1 and 4:1 families, located inside corotation for $E_j<E_{L_1}$
that support the envelope of the bar \citep{b8}.
 Some of the
characteristics of these families are plotted in Fig.~10. Moreover,
in Fig.~4b we see many real particles that stick around various
resonances both inside and outside corotation (see the various
spikes corresponding to different resonances).

\subsection{Reconstruction of the galaxy by the superposition of the main families}

The chaotic orbits close to the unstable periodic orbits
corresponding to different Jacobi constant values may form different
density distributions. However, the orbits at every Jacobi constant
support the density distribution of the real particles at the same
value of the Jacobi constant. Here we demonstrate what is the
contribution of a family, all along its characteristic, to the
global morphology of the galaxy. This is done in Fig.~16 for the 3:1
family. In Fig.~16a we plot 100 periodic orbits (black points) of
the 3:1 family equally sampled along its whole characteristic
together with the density contours of the galaxy (black lines).
These orbits are located inside corotation. In Fig.~16b we plot the
density distribution of 20000 orbits having initial conditions close
and along the periodic orbits of Fig.~16a integrated for 3-4 bar
rotations. Although the initial conditions are all inside corotation
supporting the bar shape, after a short transient period of
stickiness along the periodic orbits they are diffused outside
corotation and form a spiral structure which is slightly more closed
than that of the galaxy (Fig.~16b).

Figure 17 is similar to Fig.~16 but for the 2:1 (or $x1$) family.
Note that the superposition of the periodic orbits do not cover the
inner parts of the bar since we have chosen only the unstable
periodic orbits along the characteristic of the 2:1 family. In
Fig.~17b we see that the orbits starting close to the periodic
orbits of Fig.~17a form a spiral structure fitting well the inner
part of the spiral structure of the galaxy (black curves).

\begin{figure*}
\centering
\includegraphics[width=12.0cm]
{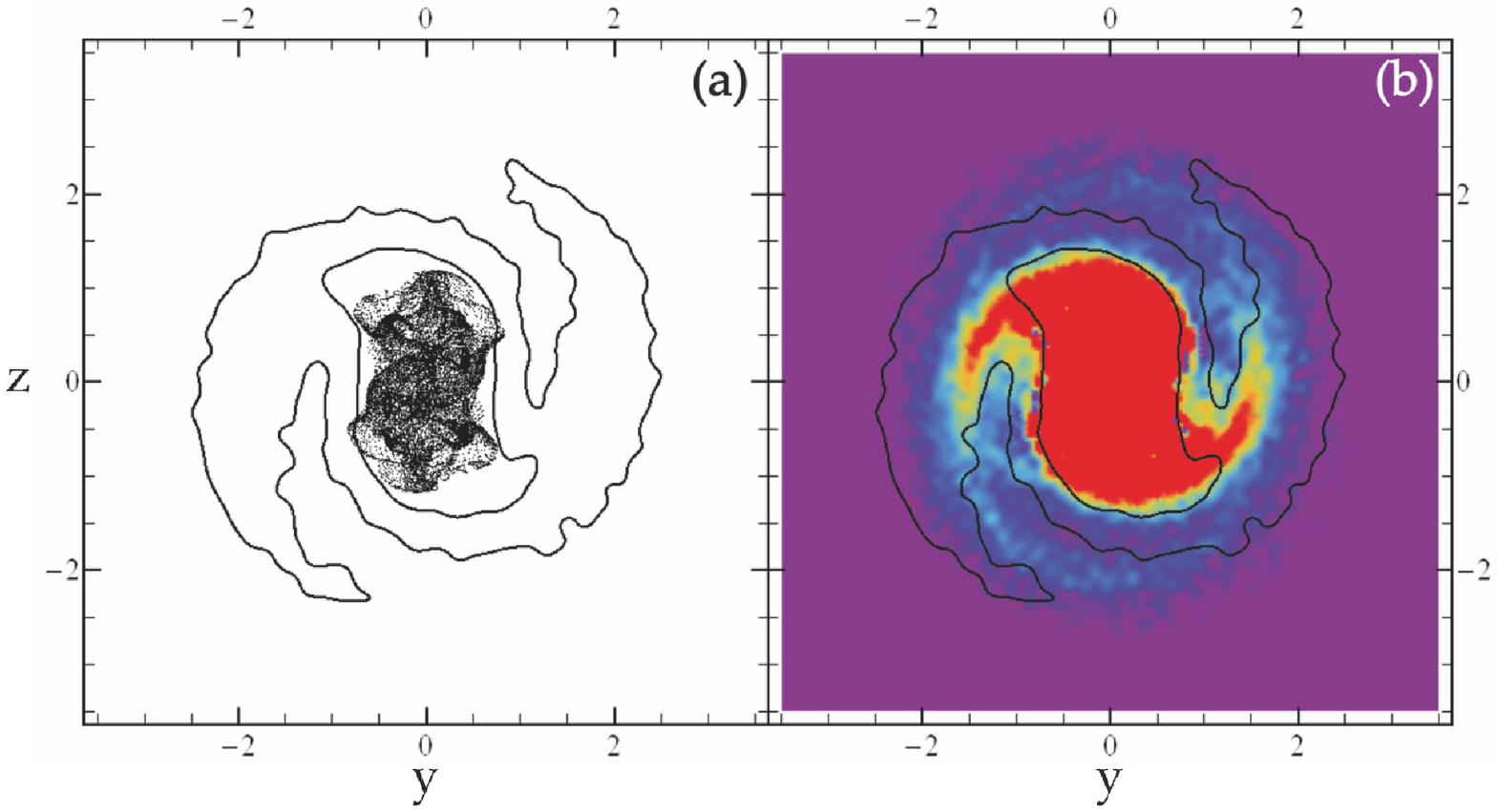} \caption{\textbf{(a)} The projection on the $y-z$ plane
of 100 periodic orbits (black points) of the 3:1 family equally
sampled along its whole characteristic together with the density
contours of the galaxy (black lines). \textbf{(b)} the density
distribution on the $y-z$ plane of 20000 orbits with initial
conditions close and along the periodic orbits shown in \textbf{(a)}
after integration for 3-4 bar rotations.} \label{fig16}
\end{figure*}

\begin{figure*}
\centering
\includegraphics[width=12.0cm]
{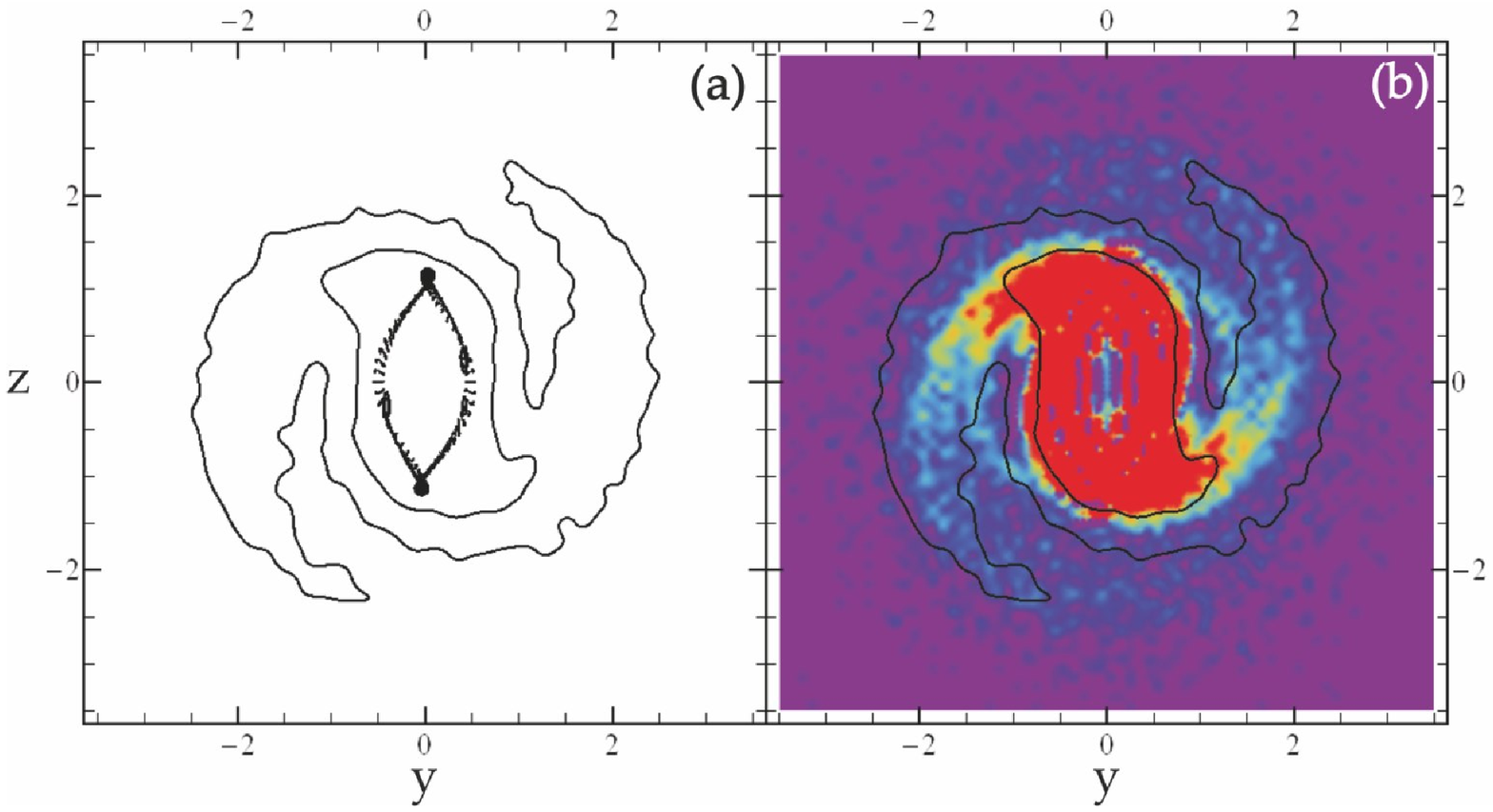} \caption{Similar to Fig.~16 but for the unstable part of
the 2:1 family.} \label{fig17}
\end{figure*}

Finally, Fig.~18 provides the same information as Figs.~16,17 but
for the $-1:1$ family. It is evident that the density maxima of the
density distribution of these orbits support the outer parts of the
spiral structure of the galaxy. The same picture is also true for
the other families outside corotation that have $q<0$ (e.g. the -2:1
family).

In Fig.~19 we plot the superimposed density distributions of the
orbits shown in Figs.~16b,17b,18b (color scale) together with the
high density contours of the galaxy (black curves). We observe that
the galaxy is well reconstructed by this set of orbits. This figure
clearly underlines the importance of the unstable periodic orbits
and of the chaotic orbits associated with them in forming the spiral
structure.

\begin{figure*}
\centering
\includegraphics[width=12.0cm]
{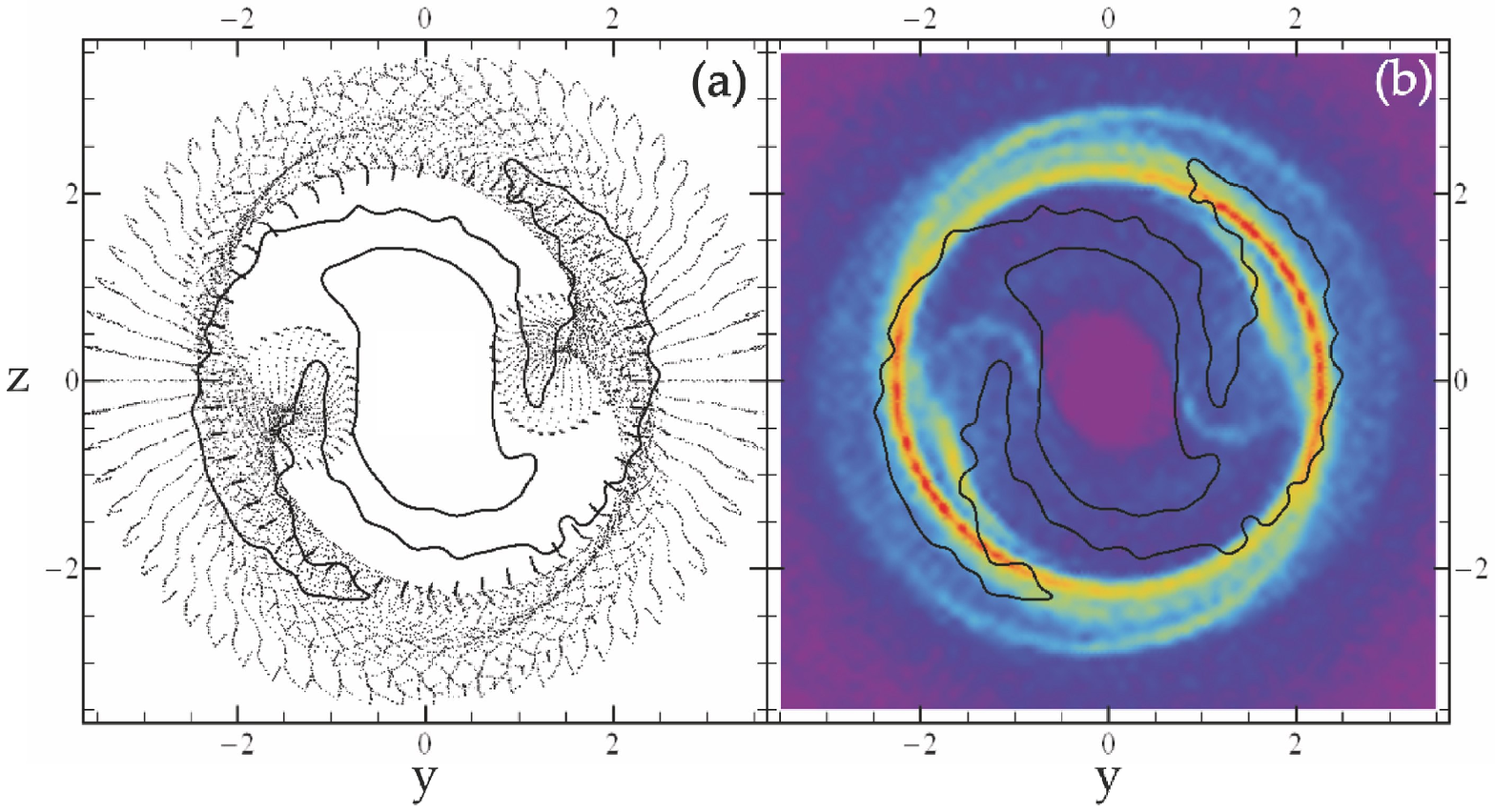} \caption{Similar to Fig.~16 but for the $-1:1$ family. }
\label{fig18}
\end{figure*}

\subsection{2D Asymptotic Orbits}

The main results of the above study, for the 3D orbits along
specific important resonances of the system can be also acquired if
we study the space distribution of the pericentres or apocentres of
the asymptotic orbits in a 2D approximation of the system ($x=0,
\dot{x}=0$ in Eq.~\ref{ejveff}). Namely, if we consider 2D unstable
periodic orbits (the most important for every energy level) and take
initial conditions along their unstable asymptotic curves, the
integration of these asymptotic orbits can also reveal the
morphological features of every energy level.
\begin{figure}
\centering
\includegraphics[width=8.5cm]
{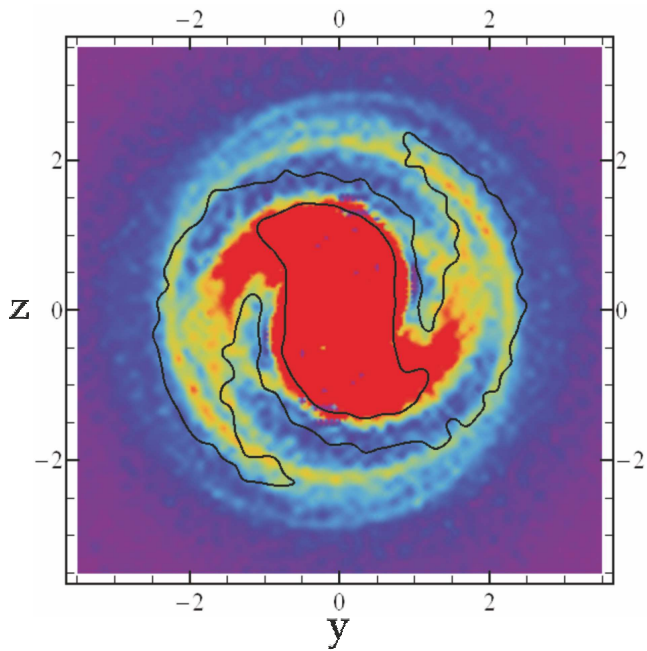} \caption{The superposition of all the distributions
shown in Figs.~16b,17b,18b superimposed with the density contours of
the galaxy (black curves). All the main morphological features of
the galaxy are reconstructed by the superposition of these orbits.}
\label{fig19}
\end{figure}
In Fig.~20a, 20b we give an example of the projection on the
configuration space of the apocentric and pericentric intersections
(black points), respectively, of the 2D manifold of the $-2:1$
unstable periodic orbit corresponding to $E_j=-1135000<E_j(L_1)$
where the areas inside and outside corotation cannot communicate.
Note that the phase space in the area outside corotation, in the 2D
approximation of the system, presents no tori or cantori with small
holes, but only a chaotic sea with escapes to infinity. In each
panel of Fig.~20 we have superimposed the density distribution (in
color scale) of the real particles at the same Jacobi constant
value. We observe that the apocentres do not coincide with the
maxima of the density while the pericentres trace well the areas of
the density maxima, but for slightly smaller radii. This is
compatible to the information provided by Fig.~12 according to which
for Jacobi constant values below $E_j(L_1)$ and outside corotation
the pericentres reveal the density maxima since they lie near low
values of the plane velocity.

Figure 21 is similar to Fig.~20 but for a different Jacobi constant
$E_j$=-1120000, which is just above $E_j(L_1)$, and therefore the
area inside corotation has just communicated with the area outside
corotation. In this energy level and in the 2D approximation there
exist unstable periodic orbits of the $PL_1$, $PL_2$ and $-1:1$
families. In Fig.~21a,b we plot the apocentres and the pericentres,
respectively, of all the asymptotic orbits (having initial
conditions on the unstable asymptotic curves) of all the periodic
orbits mentioned above, superimposed on the density distribution of
the real particles in the specific Jacobi constant value. In this
case we see that the pericentres together with some parts of the
apocentres (mostly of the $PL_1$, $PL_2$ families) trace well the
density maxima of the density distribution. This is compatible to
the information provided by Fig.~13 according to which the
pericentres and the apocentres for Jacobi constant values above
$E_j(L_1)$ reveal the density maxima when they lie near low values
of the plane velocity.

\begin{figure*}
\centering
\includegraphics[width=12.0cm]
{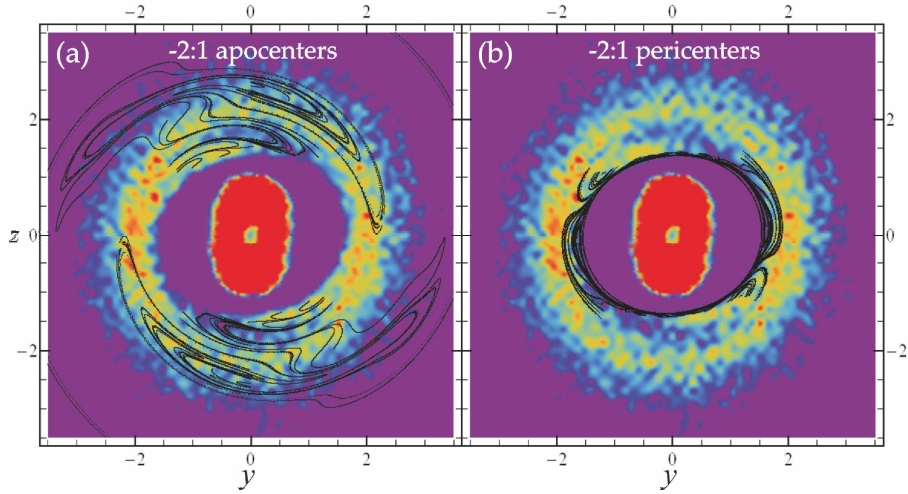} \caption{\textbf{(a), (b)} The projection on the
configuration space of the apocentric and pericentric intersections
(black points), respectively, of the 2D manifold of the $-2:1$
unstable periodic orbit corresponding to $E_j=-1135000<E_j(L_1)$
where the areas inside and outside corotation cannot communicate
superimposed on the density distribution of the real particles of
this Jacobi constant value. } \label{fig20}
\end{figure*}

\begin{figure*}
\centering
\includegraphics[width=12.0cm]
{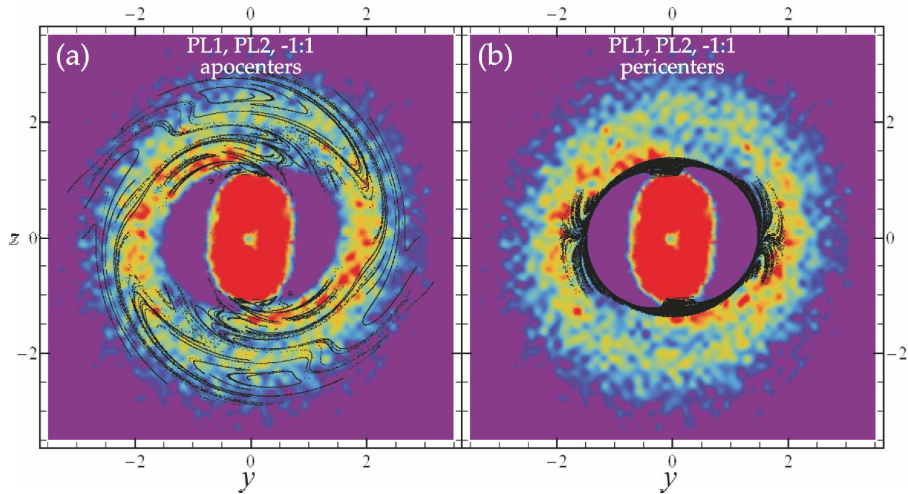} \caption{Similar to Fig.~20 but for the manifolds of the
$PL_{1,2}$ and $-1:1$ families and for $E_j=-1120000$ just above
$E_j(L_1)$.} \label{fig21}
\end{figure*}

\section{Rays, Bell-type curves and Escaping orbits}

The stickiness effect along the asymptotic manifolds of the unstable
periodic orbits can last for a number of dynamical times reinforcing
the spiral structure, as it has been described in the previous
sections, but then the chaotic orbits can escape to very large
distances and finally they may escape to infinity. This diffusion is
relatively slow and its time scale can be compared to the age of the
Universe. In this section we study the way these chaotic orbits
escape to infinity.
\begin{figure*}
\centering
\includegraphics[width=\textwidth]
{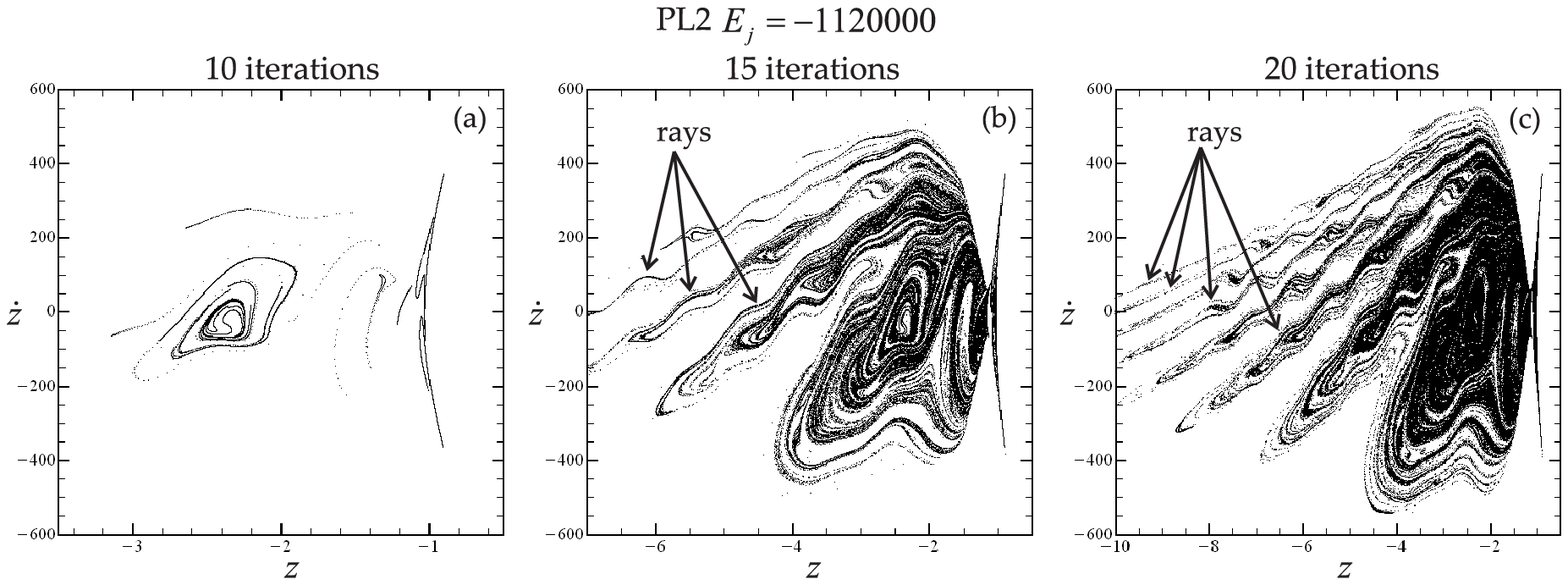} \caption{The asymptotic curves of the unstable periodic
orbit $PL_2$ in the phase space ($z, \dot{z}$) (for $y=0$ and
$\dot{y}>0$) for the 2D approximation of our model and for
$E_j=-1120000$ close to $E_j(L_1)$ after \textbf{(a)} 10 iterations
\textbf{(b)} 15 iterations and \textbf{(c)} 20 iterations of a small
initial segment along the unstable directions of the manifold.}
\label{fig22}
\end{figure*}

Figure 22 presents the unstable asymptotic curve of the $PL_2$ orbit
on a Poincar$\acute{e}$ surface of section $z,\dot{z}$ for $y=0$ and
$\dot{y}>0$ of the 2D approximation of the system and a Jacobi
constant close to $E_j(L_1)$. We have taken a small initial segment
of 50000 points and of length $10^{-6}$ from the unstable periodic
orbit and along the direction of the unstable asymptotic curve, and
we have integrated all these points for 10 iterations (Fig.~22a) 15
iterations (Fig.~22b) and 20 iterations (Fig.~22c). We observe the
formation of characteristic rays along the asymptotic curves when
integrated for a long time.

In Fig.~23 we see the picture of the same Poincar$\acute{e}$ surface
of section after the integration of 100 initial conditions of test
particles for 200 iterations in the 2D approximation of the system
(black points). These points are concentrated close to the unstable
asymptotic curves of the families $PL_1$, $PL_2$.
 In Fig.~23 we also plot the curves corresponding to zero
inertial energy $E=0$ (thick black curves). The analytic relation
$\dot{z}=f(z)$ (for $y=0$) describing these curves is found as
follows: By subtracting Eqs.~\ref{ejveff} and \ref{einert} for $E=0$
and $\dot{x}=x=y=0$ we find
\begin{equation}
\Omega_p \dot{y}z+\Omega^2_p z^2=-E_j
\end{equation}
where
\begin{equation}
\dot{y}=\pm\sqrt{2E_j+\Omega^2_p z^2-2V(0,0,z)-\dot{z}^2}
\end{equation}
Hence
\begin{equation}
\dot{z}^2=-2V(0,0,z)-\frac{E^2_j}{\Omega^2_p z^2}
\end{equation}

This last equation gives two symmetric curves, one with $\dot{z}<0$
and the other with $\dot{z}>0$, shown by thick black curves in
Fig.~23. Orbits starting below the lower thick black curve escape to
infinity without intersecting the $E=0$ curve. An example is given
by the blue dots, numbered 1,2,3,..., which correspond to the
successive iterations of an initial condition near the point $1$,
having $E>0$. These iterations are located along rays. The
corresponding orbit escapes from the galaxy following a spiral path,
in the rotating configuration space (see inserted orbit in Fig.~23).
On the other hand, if an initial condition has $E<0$, then the
successive iterations (red dots) lie on rays again, but between the
two $E=0$ curves, and at the same time they form ``bell-type''
curves (Contopoulos and Patsis 2006). Different bell-type curves
correspond to different values of the energy $E$. The value of $E$
along an orbit remains roughly constant far from the galaxy (left
part of Fig.~23) forming a bell-type curve, but it changes abruptly
whenever the orbit comes close to the main body of the galaxy (small
absolute values of $z$) and then it follows a new bell-type curve.
The changes of $E$ are irregular, due to the chaotic character of
the orbit. After several changes of its value, $E$ becomes positive,
and then the orbit escapes to infinity following a spiral-like path.
Note that this path corresponds to an open hyperbolic curve in an
inertial frame of reference.

\begin{figure}
\centering
\includegraphics[width=9.cm]
{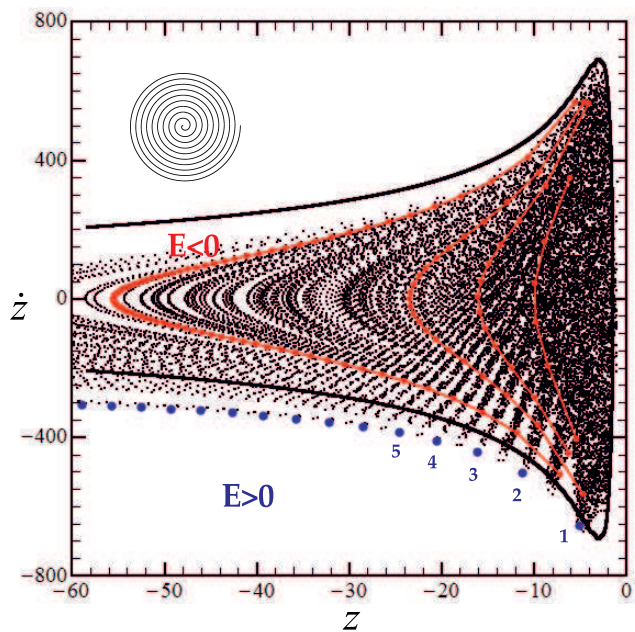} \caption{The phase space $z$, $\dot{z}$ (for $y=0$ and
$\dot{y}>0$) of a 2-D approximation of our model for a Jacobi
constant $E_j=-1120000> E_j(L_1)$, where the area inside corotation
can communicate with the area outside corotation. The thick black
curves correspond to zero inertial energy. An initial condition near
the point $1$ and inertial energy $E>0$ has successive iterations
(blue dots) along rays and escape to infinity following a spiral
like orbit (inserted orbit). On the other hand if we take initial
conditions having $E<0$ the successive iterations (red dots) lie
along rays and along bell type curves at the same time.}
\label{fig23}
\end{figure}

\begin{figure}
\centering
\includegraphics[width=8.5cm]
{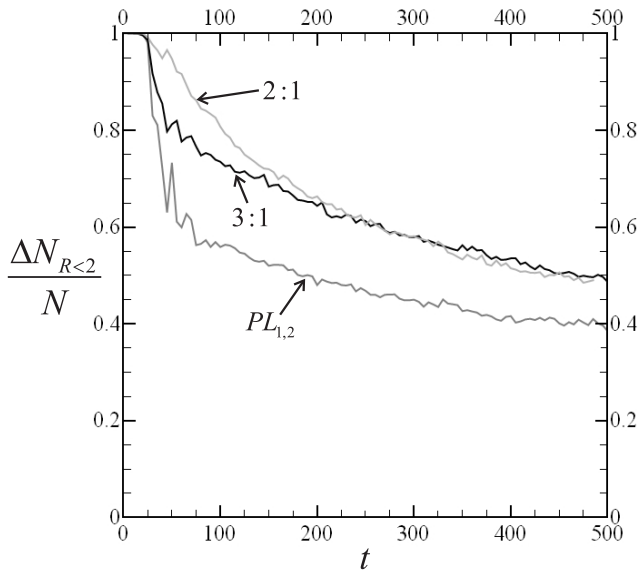} \caption{The percentage of the orbits starting close to
the 3:1, 2:1, $PL_{1,2}$ unstable periodic orbits (see Figs.~13-15)
that stay located inside $R=2r_{hm}$ as a function of time in
$T_{hmct}$.} \label{fig24}
\end{figure}

Finally, Fig.~24 presents the fraction of the particles starting
close to the 3:1, 2:1, $PL_{1,2}$ periodic orbits (see Fig.~13-15)
that stay located inside a radius of $2r_{hm}$ as a function of time
in $T_{hmct}$. This figure testifies the phenomenon of stickiness
along the manifolds originating from the unstable periodic orbits
and gives a measure of the chaotic diffusion of the system. We
observe two different rates of diffusion: \textbf{(a)} the first one
lasts for 50 to 100 $T_{hmct}$ where about the 20\%, 25\% and 45\%
of the orbits around the 2:1, 3:1 and $PL_{1,2}$ periodic orbits
respectively have been diffused outwards and \textbf{(b)} a quite
slower diffusion starting after $\approx100 T_{hmct}$. In general,
orbits with initial conditions closer to lower order $n$ of
resonance $n:1$ present a slower diffusion rate. During the first
diffusion period, the spiral structure is viable, because the flow
of chaotic material coming from inside corotation stay confined
close and along the asymptotic manifolds of the unstable periodic
orbits and support the spiral structure of the galaxy. This time
period corresponds to about 10 rotations of the bar. After that and
during the second period of slow diffusion where chaotic orbits move
outwards, spiral structure is not clearly observed in the system.

\section{CONCLUSIONS}

In this paper we study the main factors that affect the formation
and the longevity of the stellar spiral arms of barred spiral
galaxies. We examine the role of the apocentres, of the pericentres
and of the velocity minima as regards the morphological features of
a galaxy. We emphasize the role of the asymptotic orbits, emanating
from the unstable periodic orbits, in forming different segments of
the spiral arms corresponding to different values of the Jacobi
constant inside and outside corotation. We show that there is
stickiness along these asymptotic orbits which affects the longevity
of the spiral arms, which finally fade away, but after more than 10
bar rotations. Below we present and discuss our main results.

\begin{description}
    \item[\textbf{(a)}] An objective way to find the maxima of density is by finding
    the loci
of the velocity minima on the rotation plane, since the particles
spend most of their time there. These loci are:
\begin{enumerate}
    \item The apocentres for the particles with Jacobi constants
below $E_{L_1}$ that are allowed to move only inside corotation.
These apocentres, of regular or chaotic orbits, support the shape of
the bar. The regular orbits form the main body of the bar, while the
chaotic orbits form an envelope of the bar.

    \item The pericentres for the particles with Jacobi constants
below $E_{L_1}$ that are allowed to move only outside corotation.
These pericentres support the outer parts of the spiral structure of
the galaxy.

    \item An elliptical locus corresponding to the local maximum of
the effective potential $V_{eff}$ for the particles with Jacobi
constant $E_j>E_j(L_4)$ that are allowed to move both inside and
outside corotation. This locus is only partly realized in the
distribution of the real $N$-body particles, because an important
percentage of particles with $E_j>E_j(L_4)$ have minimum velocities
$v_{min}$ that do not differ significantly from the mean velocity
$<v_{yz}>$ all along their orbits. Therefore, this locus is not
reflected in the global configuration of the particles although it
is reflected in the sub-population corresponding to low $v_{min}$
values compared to $<v_{yz}>$.
\end{enumerate}
For Jacobi constants $E_j(L_4)>E_j>E_j(L_1)$ some of the loci of
pericentres and apocentres that are close to low values of the
velocity are correlated to the inner parts of the spiral structure.

    \item[\textbf{(b)}] For each value of the Jacobi constant
there is a set of stable and unstable periodic orbits. It is well
known that the domain of phase space associated with the stable
periodic orbits supports features similar to the morphologies of
these orbits.

 \item On the other hand in the present study we emphasize the role of the
unstable periodic orbits and the manifolds emanating from them.
Although the unstable periodic orbits themselves do not support the
spiral structure, the asymptotic orbits that start near the unstable
periodic orbits support the spiral structure. As long as there is a
particle population near the unstable periodic orbits of the main
families of orbits (e.g. -1:1, -2:1, $PL_1$, $PL_2$, $PL_4$, $PL_5$,
4:1, 3:1 and 2:1) the spiral structure remains clearly visible. We
show that the superposition of orbits with initial conditions close
to the periodic orbits of only three main families (-1:1, 2:1, 3:1)
all along their characteristics are able to reconstruct all the main
morphological features of the galaxy.

    \item[\textbf{(c)}] For Jacobi constants above $E_j(L_1)$ chaotic
diffusion leads to escapes. The diffusion is fast (for all the
families) for a time interval lasting for about 10 rotations.
Afterwards, the diffusion rate slows down but the spiral structure
is marginally discernible. For the families inside and close to
corotation (e.g. 2:1, 3:1, 4:1, $PL_1$, $PL_2$) the diffusion rate
is smaller for orbits closer to the center. On the other hand,
chaotic orbits having initial conditions outside corotation present
stickiness at the resonances $-1:1$ and $-2:1$ and support the outer
parts of the spiral structure.

The stickiness of the chaotic orbits along the unstable asymptotic
curves, is called ``stickiness in chaos'' \citep{b11}, as it is not
due to islands of stability or cantori. The shape of these
asymptotic curves in the phase space ("rays") corresponds to the
shape of the spiral structure in the configuration space.

 \item In our model, the phase space outside corotation is not bounded.
Therefore, chaotic orbits above and outside corotation will finally
acquire positive energy and escape from the system, along spiral
orbits (in the rotating frame of reference). The successive
iterations of chaotic orbits with negative inertial energy, in the
phase space, lie on bell-type curves, each corresponding to the same
energy. However, the energy changes when a star approaches the main
body of the galaxy. Once a chaotic orbit acquires positive energy it
has successive iterations that lie on rays that extend to infinity.
This escape will happen, in general, after times considerably longer
than the age of the Universe.
\end{description}

Finally, we note that our model is an adiabatic approximation of a
real $N$-body system. Such a study is capable of providing the main
trends about the dynamical mechanisms of the formation and the
longevity of the various structures. However, the dynamical behavior
of the real system is definitely more complicated since it shows an
evolution (decrease) of the pattern speed (although this decrease is
rather small). Another issue has to do with the evolution of the
density and consequently of the potential. In our ``frozen'' model
the amplitude of the spiral perturbation in the density decreases
with time. The change of these two parameters affects the phase
space structure. This implies a different network of unstable
periodic orbits and also different asymptotic orbits with different
stickiness around them. An extended study of the real evolving
system requires the use of dynamical methods that take into account
the evolution of both the pattern speed and the potential. We have
proceeded in this direction and we will present our results in a
forthcoming paper.

\section*{Acknowledgments}
This work has been partially supported by the Research Committee of
the Academy of Athens through the project 200/739.

\label{lastpage}

\end{document}